\documentclass[doublecol]{epl2} 
\usepackage{epsfig}
\usepackage{amsmath}
\usepackage{amsfonts}
\usepackage[svgnames]{xcolor}
\usepackage[colorlinks=true, linkcolor=Maroon, urlcolor=Blue]{hyperref} 
\usepackage{color}
\usepackage{microtype}
\DisableLigatures{encoding = *, family = *}
\usepackage[figuresleft]{rotating}
\usepackage{caption}
\usepackage{relsize}
\usepackage{subfigure}
\usepackage{graphicx}
\hypersetup{
  colorlinks,
  citecolor=MidnightBlue,
  linkcolor=DarkRed,
  urlcolor=Blue}

\title{Shot Noise as a probe for the  pairing symmetry of Iron pnictide superconductors}
\author{ Colin Benjamin\inst{1} and Tusaradri Mohapatra\inst{1}}
\institute{ School of Physical Sciences, National Institute of Science Education \& Research, HBNI, Jatni-752050,\ India                    
  \inst{1} }
\abstract{ One of the outstanding problems in Iron pnictide research is the unambiguous detection of its pairing symmetry. The most probable candidates are the two-band $s_{++}$ and sign reversed $s_{\pm}$ wave pairing. In this work the Andreev conductance and shot noise are used as a probe for the pairing symmetry of Iron pnictide superconductors. Clear differences emerge in both the zero bias differential conductance and the shot noise in the tunneling limit for the two cases enabling an effective distinction between the two.   }
 \begin{document}
\maketitle
\section{Introduction} Andreev conductance and shot noise across a Metal-Superconductor\cite{been} or Ferromagnet-Superconductor\cite{buzdin} have been subjects of extensive research in the past two-three decades. The main purpose of research in such setups is to probe their applications in tasks ranging from detection of pairing symmetry of superconductors\cite{dwave} to quantum information processing\cite{noise}. In this respect while conductance calculations have been used extensively to probe the pairing symmetry, there is no record of the use of shot noise in such tasks.  Shot noise has been used to measure the unit of transferred charge in fractional quantum hall experiments, in distinguishing particles from waves and as entanglement detector too\cite{noise}. In contrast, probably for the first time, in this manuscript shot noise will be used to detect pairing symmetry of an Iron pnictide superconductor.

The aim of this work is to propose differential conductance and shot noise as a possible discriminator between the two possible $s_{++}$ and $s_{\pm}$ pairing symmetries of Iron based superconductors\cite{stewart}. Experimental tests like the half-flux quantum\cite{liu} have utilized Josephson coupling, between an Iron superconductor and a s-wave superconductor, and have managed to zero in on the $s_{\pm}$ pairing but doubts remain\cite{PRL120}. The spontaneous magnetic flux measured can identify the sign-reversed pairing symmetry ($s_{\pm}$) in Josephson junction with Iron-based superconductor. In a recent work, the feasibility of tuning the coupling between two bands of the Iron superconductor was discussed so as to discriminate between the two possible pairing symmetries\cite{PRL120,PRB86}.~The Josephson coupling changes from adding constructively for $s_{++}$ case to canceling destructively for $s_{\pm}$ case due to the $\pi$ phase shift. Thus due to phase sensitivity of Josephson junctions, there is almost complete cancellation of supercurrents from sign-reversed pairing symmetry in Iron pnictide Josephson junctions\cite{PRL120}. We will also exploit this property in Iron superconductors to discriminate between the two pairing symmetries via the differential conductance and shot noise.

 Two tunneling channels in Iron pnictide based junctions are due to the multiband nature of the Iron-superconducting electrode. This gives rise to complicated interference depending on the underlying pairing symmetry\cite{PRB86}. We show it is the interference of waves reflected from different pairing symmetries of Iron pnictide superconductor junctions which helps in distinguishing between them. The layout of the paper is as follows: in the next section we briefly discuss the competing pairing symmetries in Iron superconductors and how they arise, next we discuss the first of our chosen settings namely a Normal Metal-Insulator-Normal Metal-Insulator-Iron pnictide junction focussing on the wavefunctions, boundary conditions and expressions for differential conductance and shot noise. After this we discuss the second setting a Ferromagnet-Insulator-Normal Metal-Insulator-Iron pnictide junction. This is followed by a discussion on the results for both the settings. We finally conclude with a note on experimental realization of our chosen settings. 
\section{Theory of electron and hole pockets in Iron superconductors}
\begin{figure}[h]
\centering
\includegraphics[scale=0.5]{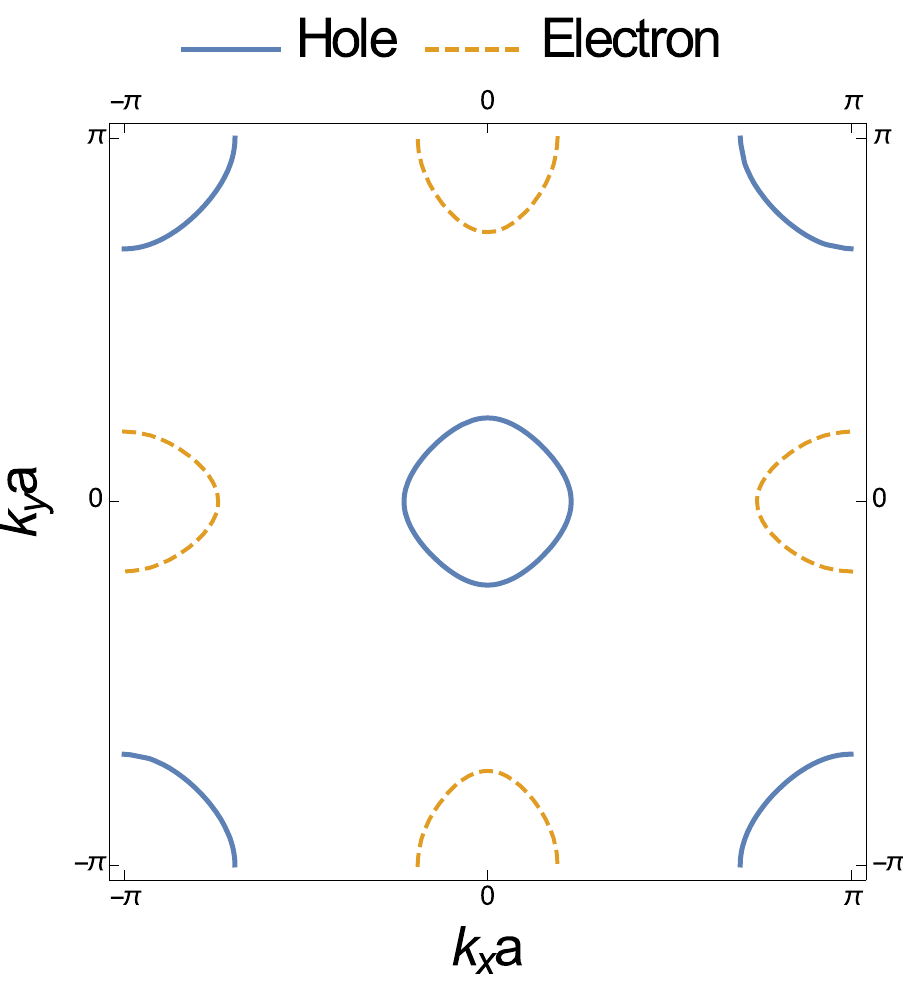}
\caption{Electron and hole packets in the brillouin zone of Iron pnictide superconductor.}
\label{fig:1}
\end{figure}

The kinetic energy term of an Iron pnictide superconductor can be derived using a tight binding model \cite{PRB220503}: 
\begin{equation}
H_{Kinetic} \!\!= \!\!\left( 
 \begin{array}{cc} 
   \varepsilon_x - \mu & \varepsilon_{xy} \\ 
   \varepsilon_{xy} & \varepsilon_y - \mu
 \end{array} 
\right),
\end{equation}
where $\varepsilon_{x} = -2 t_{1} \cos(k_xa) - 2 t_{2} \cos(k_ya) - 4 t_{3}\cos(k_xa)\cos(k_ya)$, $\varepsilon_y = -2t_{2}\cos(k_xa) - 2 t_{1}\cos(k_ya) - 4t_{3} \cos(k_xa)\cos(k_ya)$, $\varepsilon_{xy} = -4t_{4}\sin(k_xa) \sin(k_ya)$ and $\mu$ denotes the chemical potential with $a$ being the lattice constant. For the parameters $t_{1} = -1$, $t_{2} = 1.3$, $t_{3} = t_{4} = 0.85$ and $\mu= 0.45$ the FeAs (Iron pnictide) band structure is plotted in Fig.~\ref{fig:1}. The Fermi surfaces obtained by diagonalizing $H_{Kinetic}$ are plotted in the unfolded Brillouin zone, it has two electron pockets(or, electron bands) centered at (0,$\pm\pi$) and ($\pm\pi$,0) and two hole pockets(or, hole bands) centered at (0,0) and ($\pi,\pi$). If the Iron pnictide superconductor lies on the $x-y$ plane, an incident electron at the metal-superconductor interface with small $p_y$ is transmitted through the electron and hole Fermi surface pockets. In this work we follow the assumption in Ref.~\cite{linder} and consider the Andreev reflection problem as envisaged with a Fermi surface consisting of two hole and two  electron pockets or bands.
The problem can be generalized to the four pocket Fermi surface shown in Fig.~\ref{fig:1} as in Refs.~ \cite{PRB220503,RPP}. Another important point to note from Fig.~\ref{fig:1} is the translation in-variance in the $y$-direction \cite{PRL120}. The full Hamiltonian of the Iron pnictide superconductor is then a sum of the Kinetic energy term and pairing potential and can be written as:
\begin{equation}
\!\!H\!\!=\!\!H_{Kinetic}\!+\!V_{pairing}\!\!=\!\! 
\left(\begin{array}{cc}\!H_{kinetic}(k)&\!\Delta(k)\\ 
\!\Delta^*(k)&\!H_{kinetic}^*(k)\end{array}\!\right).
\end{equation}
The superconducting gap $\Delta(k)$ assumes two different values for the gap $\Delta_e$ and gap $\Delta_h$ in the electron and hole Fermi surfaces. In this work, we concentrate on two alternative scenarios for the pairing symmetry of Iron pnictide superconductor\cite{yctao} the two band s-wave case $s_{++}$ in which $\Delta_e$ and $\Delta_h$ have same sign and contrast it with the two band $s_{\pm}$-wave case for which $\Delta_e$ and $\Delta_h$ take on opposite signs.
\section{Metal-Insulator-Metal-Insulator-Iron pnictide superconductor junction}
\begin{figure}[h]
\vskip -0.25in
\includegraphics[scale=0.48]{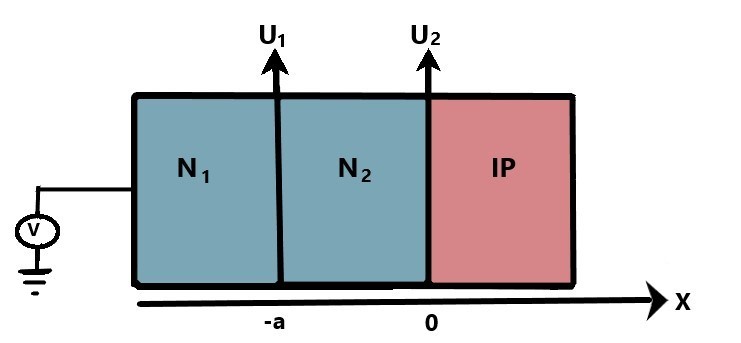}
\caption{Normal metal($N_1$)-Insulator-Normal metal($N_2$)-Insulator-Iron pnictide(Ip) superconductor junction}
\label{fig:2}
\end{figure}
In Fig. \ref{fig:2} we show the first of our chosen settings to detect the pairing symmetry of Ip superconductor. The normal metal $N_1$ is at bias voltage $V$ with respect to the metal $N_2$ and Iron pnictide superconductor which are both grounded.
Due to two non-superconducting layers, there will be multiple reflections between the two adjacent normal metals which can result in quasibound states. If there is a single layer of normal metal, there won't be any interference between the reflected electrons/holes from normal metal $N_1$ and reflected electrons/holes from normal metal $N_2$. These multiple reflections will be ofcourse dependent on the pairing symmetry phase $\Delta \phi=\phi_1-\phi_2$, where $\phi_i, i=1,2$ is the superconducting phase for band $i$, now while $\Delta \phi =0$ for $s_{++}$,  $\Delta \phi =\pi$ for $s_{\pm}$. The important point is that these multiple reflections will result in constructive/destructive interference due to the difference in phase between $s_{++}$ and $s_{\pm}$ pairing which will show up in  both conductance and shot noise calculations, see section V. The use of double barrier structure in conjunction with superconductors has been used as a probe for proximity effect\cite{volkov}, and to probe conductance oscillations\cite{ohturi} in Ferromagnet-Normal metal double barrier structure in conjunction with a s-wave superconductor.
\subsection{ Hamiltonian}
The Hamiltonian of Ip superconductor from Eq.~2 is given as below, with $\varepsilon_{k,1}$ and $\varepsilon_{k,2}$ the two electronic energy bands from Eq.~2, while $-\varepsilon_{k,1}$ and $-\varepsilon_{k,2}$ are the two hole energy bands with $\mathcal{H} \psi = E \psi$, where
{\small
\begin{equation}
\mathcal{H} = 
\left[\begin{array}{cccc}
   \varepsilon_{k,1}+U(x)  & \Delta_1(k) \Theta(x) & \alpha_{0} \delta (x) & 0  \\ 
   \Delta_1^*(k) \Theta(x)  & -\varepsilon_{k,1}-U(x) & 0 & -\alpha_{0} \delta (x)\\
   \alpha_{0} \delta (x) & 0  & \varepsilon_{k,2}+U(x)  & \Delta_2(k) \Theta(x)\\
   0 & -\alpha_{0} \delta (x)  & \Delta_2^*(k) \Theta(x)  & -\varepsilon_{k,2}-U(x)
\end{array}\right],
\end{equation}}
and $\alpha_0$ is the interband coupling strength between the two bands in Ip superconductor and $E$ defines the energy of the states. The two bands couple through the interface scattering as long as $\alpha_{0} \ne 0$ \cite{linder}. For wavefunctions and boundary conditions of $N_{1}/I/N_{2}/I/Ip$ junction, please see supplementary material.
\subsection{Conductance and shot noise in $N_{1}/I/N_{2}/I/Ip$ superconductor junction}
The well known BTK~\cite{BTK} approach to calculate the differential conductance in Normal metal-Superconductor junctions was previously extended to normal metal-Ip superconductor junction in Ref.~\cite{linder}. In this paper, we extend it to address both differential conductance and differential shot noise in both $N_{1}/I/N_{2}/I/Ip$ as well as Ferromagnet/insulator/normal metal/insulator/Iron pnictide superconductor junction as a means to detect the pairing symmetry of Ip superconductor. To calculate the currents in the normal metals one has to sum the contributions of electron incident from both bands. The net charge current induced by a voltage drop $eV$ across the junction $I_{\lambda}$ for electron incident in band $(\lambda=1,2)$ is-
\begin{eqnarray}
I_{\lambda}\!\!\!\!&=&\!\!\!\!2N(0)ev_F \mathcal{A}\!\!\sum_{\sigma=1,2}\int_{-\infty}^{\infty} \!\!\!\!\left( 1-B_{\sigma}(E) \right)\left[ \textit{f}_{0}(E-eV)-\textit{f}_{0}(E)  \right]\nonumber\\
 &+&A_{\sigma}(E)\left[\textit{f}_{0}(E) -\textit{f}_{0}(E+eV) \right] dE.
\end{eqnarray}
The incoming electrons from Ip superconductor have Fermi distribution $\textit{f}_{0}(E)$, while incoming electrons from Normal metal $N_1$ have distribution $\textit{f}_{0}(E-eV)$. In Eq.~(4) $\mathcal{A}$ is the cross sectional area of the interface, $v_F$ the Fermi velocity, $N(0)$ is the density of states at the Fermi energy $E_F$ and subscript $\sigma$ in the scattering probabilities describes whether the reflection is from band $1$ or band $2$ of Iron-based superconductor. After determining the scattering probabilities we calculate the differential conductance from Eq.~(4) as-
 \begin{equation}
 G_{\lambda}(E) \propto \!\!\!\sum_{\sigma=1,2} \!\int_{-\infty}^{\infty}  \left[\frac{\partial \textit{f}_0(E-eV)}{\partial E}\right]\!\left[1+A_{\sigma}(E)- B_{\sigma}(E)\right] dE, 
 \end{equation}
where $\lambda$ denotes incoming electron from band $\lambda=1,2$. At temperature $T=0$, Fermi function is a Heaviside theta function. Thus, we have: $-\frac{\partial \textit{f}_0(E-eV)}{\partial E} = \delta(E-eV)$. The normalized differential conductance of the system at temperature $T=0$ is then\cite{BTK,PRB51} 
\begin{equation}
G_{\lambda}(eV)\propto\frac{dI_{\lambda}/dV}{(dI/dV)}_{NM}\!\!=\!\!\!\sum_{\sigma=1,2} \!\!\left[1+A_{\sigma}(eV)-B_{\sigma}(eV)\right]/T_{NM},
\end{equation}
where $T_{NM}$ is the tunneling conductance in the normal state with Ip replaced by a normal metal. The differential conductance for two band Ip superconductor thus is given as-
\begin{equation}
G(eV)/G_0 = \frac{1}{2 T_{NM}} \sum_{\lambda=1,2} G_{\lambda}(eV),
\end{equation}
where $G_0=\frac{2 e^2}{h}$, $G_{\lambda}(eV)= 1 + A_1(eV) +A_2(eV) -B_1(eV) - B_2(eV)$ for incoming electron in band $\lambda$ and $T_{NM}$ is the transmission probability of a Normal metal-Insulator-Normal metal-Insulator-Normal metal junction.

Next, we calculate the shot noise for our junction. Shot noise is defined as the temporal fluctuation in electric current in non-equilibrium(transport) across a system. Unlike thermal noise which vanishes at zero temperature shot noise exists even at zero temperature. This is a consequence of the discreteness of charge. The general result for shot noise power\cite{PRB53} $P_{11}$ (the double subscript $11$ refers to the fact  that shot noise is current-current correlation in normal metal) across a normal metal/superconductor junction is
\begin{eqnarray}
P_{11}\!\!\!&=&\!\!\!\!\!\frac{2e^2}{h}\sum_{k,l\in 1,2;x,y,\gamma,\delta\in e,h} \int sgn(x) sgn(y) dE W_{k,\gamma;l,\delta}(1x,E) \nonumber\\
&& W_{l,\delta;k,\gamma}(1y,E) \textit{f}_{k \gamma}(E) [1-\textit{f}_{l \delta}(E)],
\end{eqnarray}
where the parameter  $W_{k,\gamma;l,\delta}(1x,E)=\delta_{1 k} \delta_{1 l}\delta_{x \gamma}\delta_{x \delta} - s^{x \gamma \dagger}_{1k}(E) s^{x \delta}_{1 l}(E)$ contains all the information about the scattering process, $s^{x \gamma}_{1 k}(E)$ represents the scattering amplitude for a particle of type $\gamma$ incident from contact $k$ which is transmitted to contact $1$ as a particle of type $x$ and $\textit{f}_{k \gamma}$ is the Fermi function for particle of type $\gamma$ in reservoir $k$. It should be noted that normal metal is contact $1$ while superconductor is contact $2$. Here $sgn(x)=+1$ for $x=e$, i.e, electron and $sgn(x)=-1$ for $x=h$, i.e., hole. Because of Andreev reflection an electron incident in contact $1$ can result in either an electron or a hole leaving contact $1$ or $2$. We can further simplify the shot noise expression by separating the electron-electron (or, hole-hole) correlations identified as $P^{AA}_{11}$ and electron-hole (or, hole-electron) correlations as $P^{AB}_{11}$. Thus, $P_{11}=P^{AA}_{11}+P^{AB}_{11}$, where $P^{AA}_{11}= \langle \Delta I_{1e} \Delta I_{1e} + \Delta I_{1h} \Delta I_{1h} \rangle$ and $P^{AB}_{11}= \langle \Delta I_{1e} \Delta I_{1h} + \Delta I_{1h} \Delta I_{1e} \rangle$. Further $P^{AA}_{11}$, $P^{AB}_{11}$ from Eq.~(8) can be written as\cite{PRB53}-
\begin{eqnarray}
P^{AA}_{11}\!\!\!\!&=&\!\!\!\!\frac{2e^2}{h} \int \sum_{x \in e,h} \{ (1-T^{xx}_{11})^2
\textit{f}_{1 x}(E) [1-\textit{f}_{1x}(E)] + \hspace{-0.18cm} \sum_{k\gamma l \delta \neq 1 x 1 x} \hspace{-0.25cm} T^{x \gamma}_{1k}(E) \nonumber \\
&& T^{\delta \gamma}_{1l}(E) W_{l,\delta;k,\gamma}(1y,E) \textit{f}_{k \gamma}(E) [1-\textit{f}_{l \delta}(E)] \} dE,\\
P^{AB}_{11}\!\!\!\!&=&\!\!\!\!\frac{2e^2}{h} \int \sum_{x \in e,h} \{ 2 T^{x \bar{x}}_{11}
\textit{f}_{1 \bar{x}}(E) [1-\textit{f}_{1 \bar{x}}(E)] + \sum_{k\gamma} s^{\bar{x} \gamma}_{1k}(E) s^{x \gamma \dagger}_{1k}(E) \nonumber \\ 
&& \textit{f}_{k\gamma}(E) \sum_{l \delta} s^{x \delta}_{1l}(E) s^{\bar{x} \delta \dagger}_{1l}(E) \textit{f}_{l \delta}(E) \} dE,
\end{eqnarray}
in Eqs.~(9-10) the scattering probabilities are related to scattering amplitudes, i.e., $T^{x \gamma}_{1k}(E)=|s^{x \gamma}_{1k}(E)|^2$ and if $x=e$ then $\bar{x}=h$. Further, at zero temperature, the term $\textit{f}_{1 \bar{x}}(E) [1-\textit{f}_{1 \bar{x}}(E)]$ vanishes, and only the second term in $P^{AA}_{11}$ and $P^{AB}_{11}$ remains. After some algebra, the shot noise power can be written as-
\begin{eqnarray}
P_{11} &=& \frac{4e^2}{h} \int^{eV}_{0} dE \{  T^{ee}_{11}(E) T^{he}_{11}(E) + T^{eh}_{11}(E) T^{hh}_{11}(E) \nonumber \\
&+& T^{ee}_{11}(E) T^{he}_{11}(E)+ T^{eh}_{11}(E) T^{ee}_{11}(E) \},\nonumber \\
&=& \frac{4e^2}{h} \int^{eV}_{0} dE \{ T^{ee}_{11}(E) (1-T^{ee}_{11}(E)) + T^{he}_{11}(E)  \nonumber \\
&& (1-T^{he}_{11}(E)) + 2 T^{ee}_{11}(E) T^{he}_{11}(E) \}.
\end{eqnarray}
 Now $T^{ee}_{11}(E)$ is the normal reflection probability $B(E)$ while $T^{he}_{11}(E)$ is the Andreev reflection probability $A(E)$. Therefore Eq.~(11) can be written in terms of $A$ and $B$ as-
\begin{eqnarray}
P_{11}\!\!&=&\!\!\frac{4e^2}{h} \int^{eV}_{0} dE \{ A(E) (1-A(E)) + B(E)(1-B(E)) \nonumber\\
&+& 2 A(E) B(E) \},
\end{eqnarray}
Eq.~(12) is the expression for shot noise power in a Normal metal-Superconductor($NS$) junction. In a $N_{1}/I/N_{2}/I/Ip$ superconductor junction due to multi-band structure of Ip superconductor, an incident electron from band $\lambda=1$ or $2$ can result in reflection of an electron and hole in bands $1$ and band $2$. Shot noise power can then be defined as $P_{11}=P_{11(1)}+P_{11(2)}$, where $P_{11 (\lambda)}$ is shot noise power for incident electron from band $\lambda=1(2)$. Shot noise power derived for $N/S$ junction, Eq.~(12), above can be extended to $N_{1}/I/N_{2}/I/Ip$ superconductor junction as follows:
 \begin{eqnarray}
    P_{11 (\lambda)}\!\!&=\!\!&\frac{4e^2}{h} \int^{eV}_{0}\!\! dE \sum_{\sigma=1,2} A_{\sigma}(E)[1- A_{\sigma}(E)]+B_{\sigma}(E)\nonumber \\
    && [1- B_{\sigma}(E)]+ 2 A_{\sigma}(E) B_{\sigma}(E),
 \end{eqnarray}
with $\lambda=1(2)$ and $P_{11 (\lambda)} = (1/e) \int S_{\lambda} dE$ where $S_{\lambda}$ being the differential shot noise for incident electron from band $\lambda$. Differential shot noise for $N_{1}/I/N_{2}/I/Ip$ superconductor junction\cite{JP,noise} is thus $S/S_0=\frac{1}{2} \sum_{\lambda=1,2} S_{\lambda},$ where $S_0 = (4e^3/h)$ and $S_{\lambda}= A_{1}(eV)(1-A_{1}(eV))-B_{1}(eV)(1-B_{1}(eV))+2A_{1}(eV)B_{1}(eV)+A_{2}(eV)(1-A_{2}(eV))-B_{2}(eV)(1-B_{2}(eV))+2 A_{2}(eV)B_{2}(eV)$ for incoming electron in band $\lambda=1,2$. One can also determine the differential Fano factor which is defined as ratio of differential shot noise to differential conductance as $F= \sum_{\lambda =1,2} S_{\lambda}/\sum_{\lambda = 1,2} G_{\lambda}$. Next we study the differential conductance and shot noise in a Ferromagnet-Insulator-Normal Metal-Insulator-Iron pnictide superconductor($FM/I/NM/I/Ip$) junction.
\section{$FM/I/NM/I/Ip$ junction}
The  $FM/I/NM/I/Ip$ superconductor setting is shown in Fig.~1 of supplementary material it's basically same as the $N_{1}/I/N_{2/I/Ip}$ junction with $N_1$ layer replaced by Ferromagnet. For wavefunctions and boundary conditions of $FM/I/NM/I/Ip$ junction, please see supplementary material.
\subsection{Conductance and shot noise for $FM/I/NM/I/Ip$ superconductor junction}
The differential conductance for two-band Ip superconductor normalized \cite{linder} by $G_0=2 e^2/h$ within the BTK formalism for $FM/I/NM/I/Ip$ superconductor junction can be calculated similarly to that for $N_{1}/I/N_{2}/I/Ip$ junction, and is given as:
\begin{equation}
G(eV)/G_0 = \frac{1}{2 T_{FM}} \sum_{\lambda=1,2} G_{\lambda}(eV).
\end{equation}
where $G_0=(2 e^2)/h, G_{\lambda}(eV)=1 + A_1(eV) + A_2(eV) - B_1(eV) - B_2(eV)$ for incoming spin up electron in band $\lambda$ and $T_{FM}$ being the transmission probability through a $FM/I/NM/I/NM$ junction. For $FM/I/NM/I/Ip$ superconductor junction the differential shot noise too can be calculated as done before for $N_{1}/I/N_{2}/I/Ip$ junction by generalizing the Andreev shot noise across a Normal metal-Superconductor junction\cite{JP}, as follows: $S/S_0 = \frac{1}{2} \sum_{\lambda=1,2} S_{\lambda},$ where $ S_{\lambda} = A_1(eV) (1-A_1(eV)) + B_1(eV)(1 - B_1(eV))+ 2 A_1(eV) B_1(eV) + A_2(eV) (1 - A_2(eV)) + B_2(eV)(1 - B_2(eV)) + 2 A_2(eV) B_2(eV) $ and $S_0 = (4e^3)/h$ , with $S_{\lambda}$ being the differential shot noise for spin up electron incident in band $\lambda$. The differential Fano factor is defined as ratio of differential shot noise to differential conductance, i.e., $F= \sum_{\lambda=1,2} S_{\lambda}/\sum_{\lambda=1,2} G_{\lambda}$. 
\section{Results and Discussion} 
In this section, for $s_{++}$ and $s_{\pm}$ pairing in Ip superconductor, we calculate the differential conductance, differential shot noise and differential Fano factor for the superconducting gap ratio $\beta = \Delta_2/\Delta_1$ as $1.5$ and for barrier strengths $z_{1}=z_{2}=z$, first for $N_1/I/N_2/I/Ip$ junction and then for $FM/I/NM/I/Ip$ junction.
\section{$N_1/I/N_2/I/Ip$ superconductor junction}
\subsection{Differential conductance}
\begin{figure}[h]
\hspace{-0.6cm}  \subfigure[]
               { \includegraphics[scale=0.285]{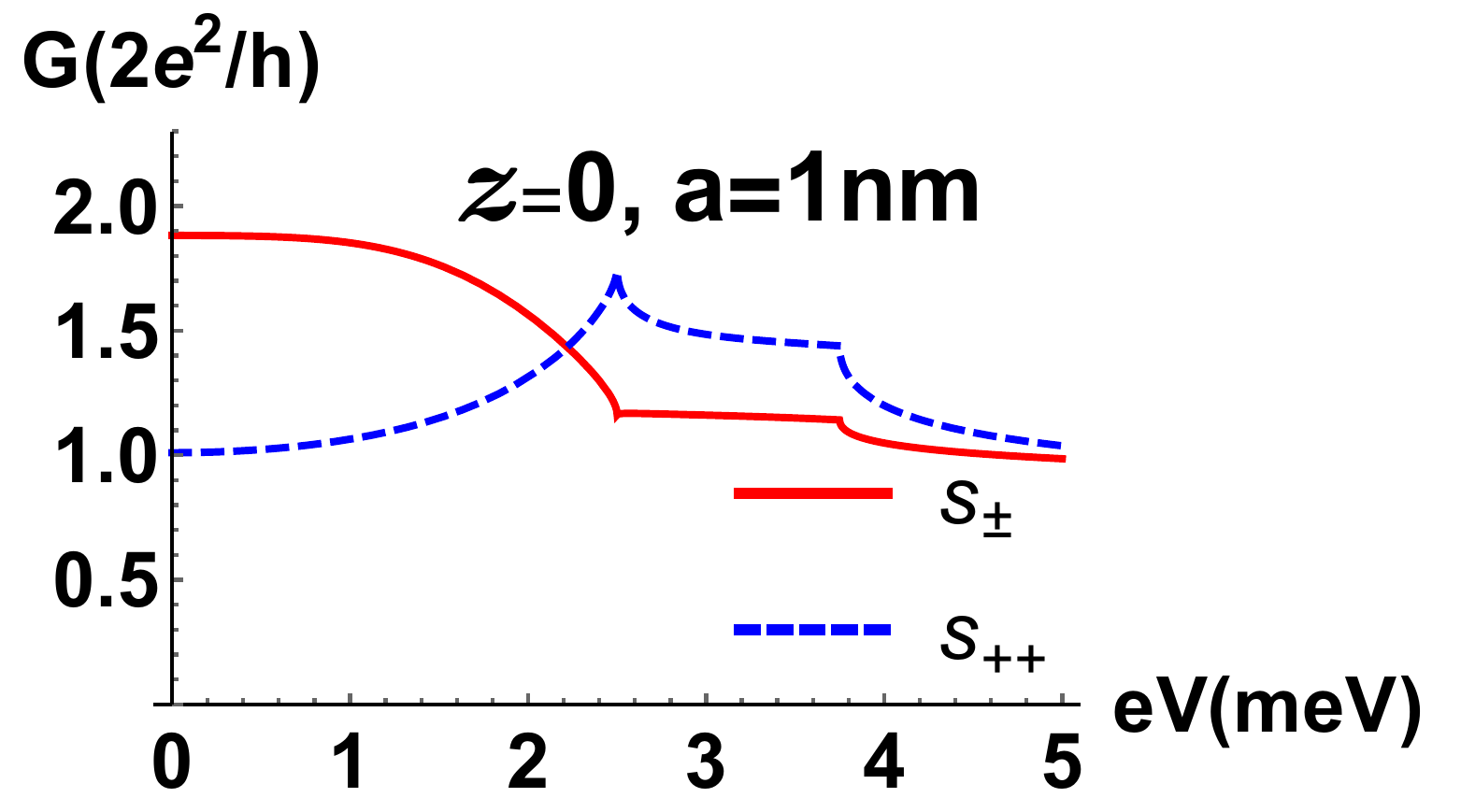} }  
  \hspace{-0.5cm}      \subfigure[]
                {\includegraphics[scale=0.285]{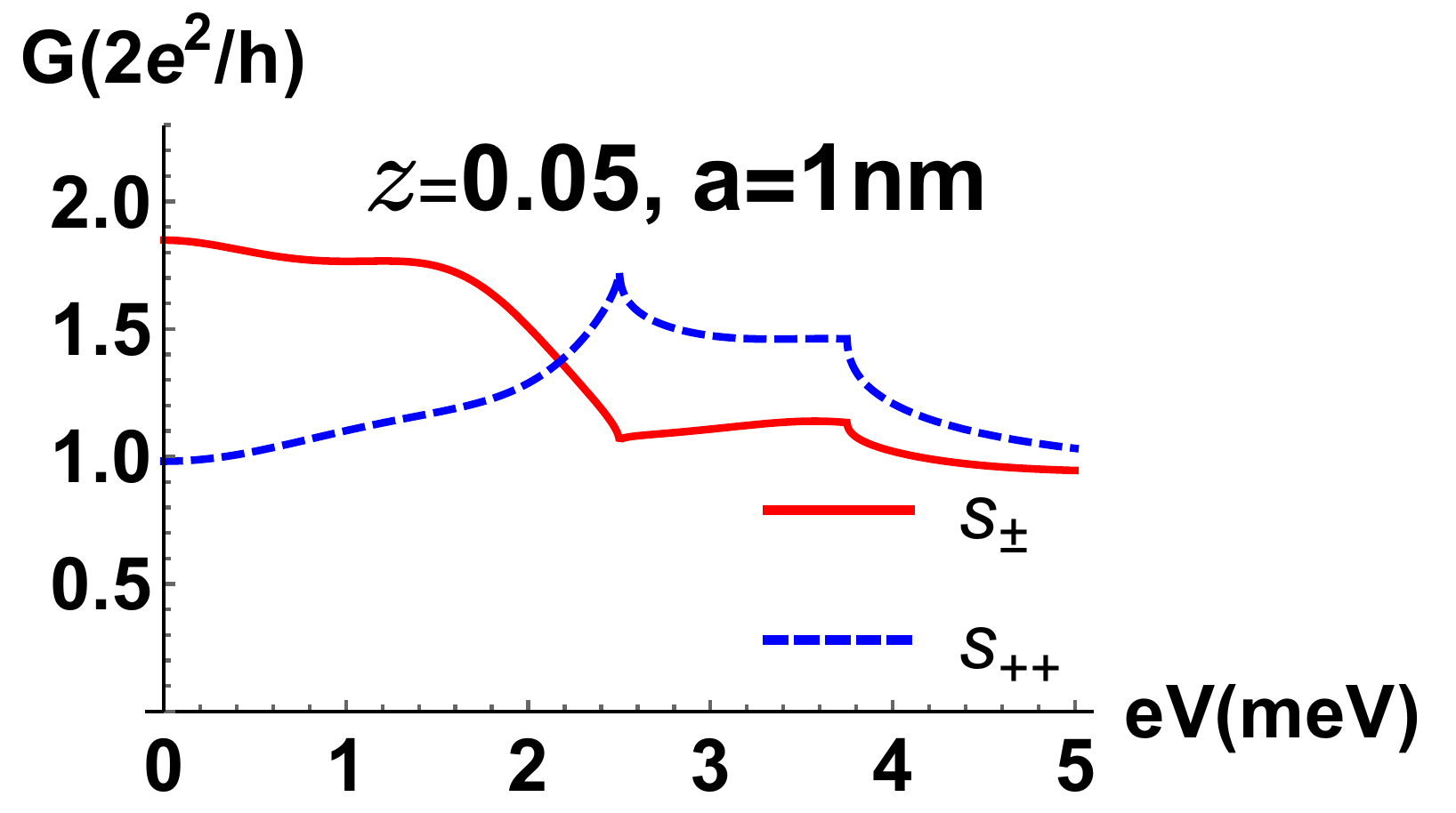} }     
  \hspace{-0.5cm}      \subfigure[]
                  { \includegraphics[scale=0.285]{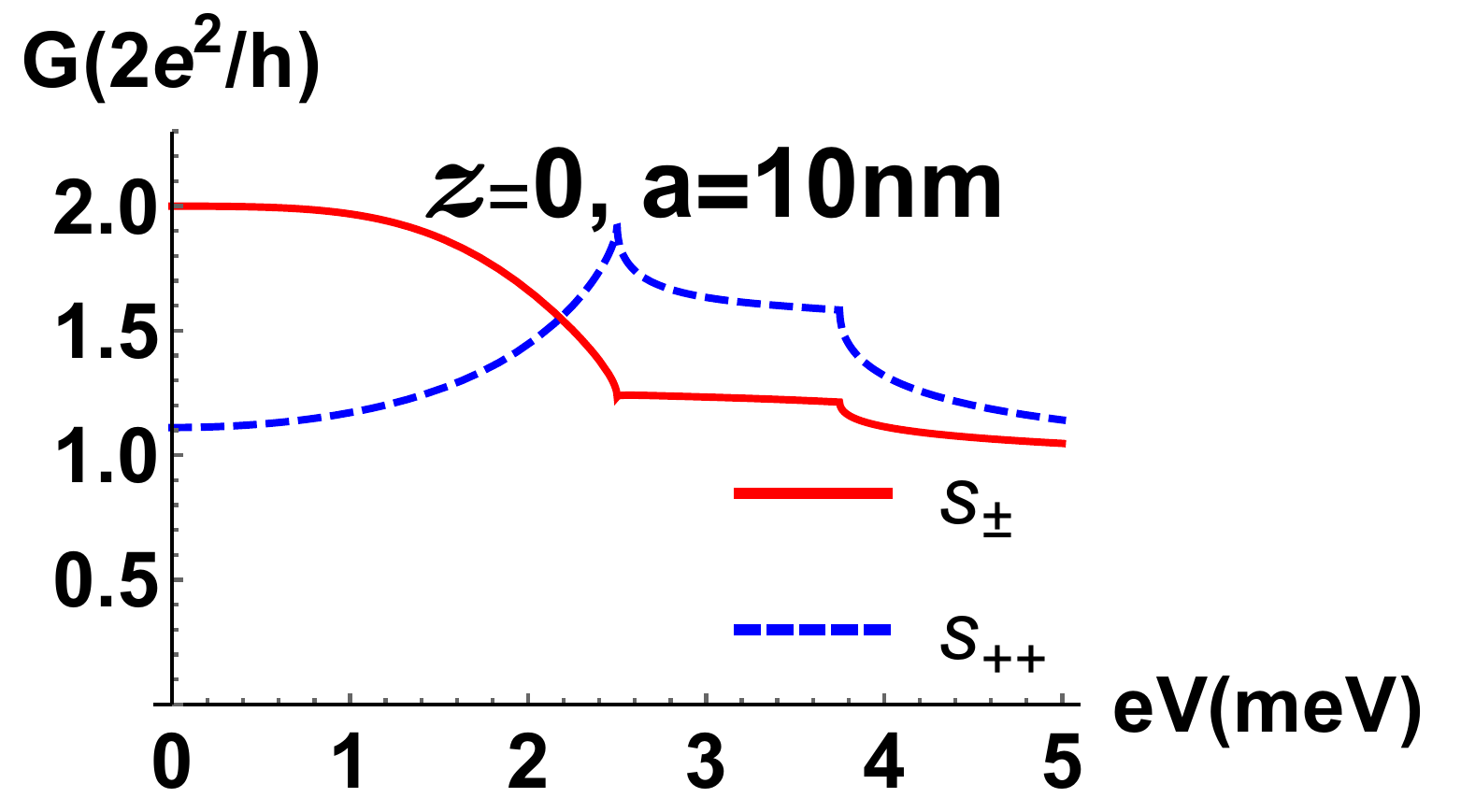}}   
        \caption{Normalized differential conductance for a $N_1/I/N_2/I/Ip$ superconductor junction vs bias voltage V(meV) with $\Delta_2$=3.75meV, $\Delta_1$=2.5meV, $E_F$=3.8eV and $\alpha$=1 for (a) $a=1nm$ and $z_{1}=z_{2}=z=0$, (b) $a=1nm$ and $z=0.05$, (c) $a=10nm$ and $z=0$. }
\end{figure}
As a first application of our model, we plot the differential conductance for a $N_1/I/N_2/I/Ip$ superconductor junction vs the bias voltage to illustrate the influence of barrier strengths and also focus on the zero bias limit for both $s_{++}$ and $s_{\pm}$ pairing symmetries. In Fig.~3(a) and Fig.~3(c), the differential conductance in $N_1/I/N_2/I/Ip$ junction for s$_\pm$ pairing shows zero bias conductance peak(ZBCP) while s$_{++}$ pairing shows a dip at zero bias regardless of any small change in thickness of the intermediate layer (for example $a=1nm$ and $a=10nm$). s$_\pm$ pairing shows ZBCP while s$_{++}$ pairing shows a dip at zero bias for transparent barrier strength and also for any small change in barrier strength($z$) as shown in Fig.~4(b).  
\subsection{Differential shot noise and differential Fano factor}
\begin{figure}[h]\hspace{-0.5cm}
        \subfigure[]
         {\includegraphics[scale=0.28]{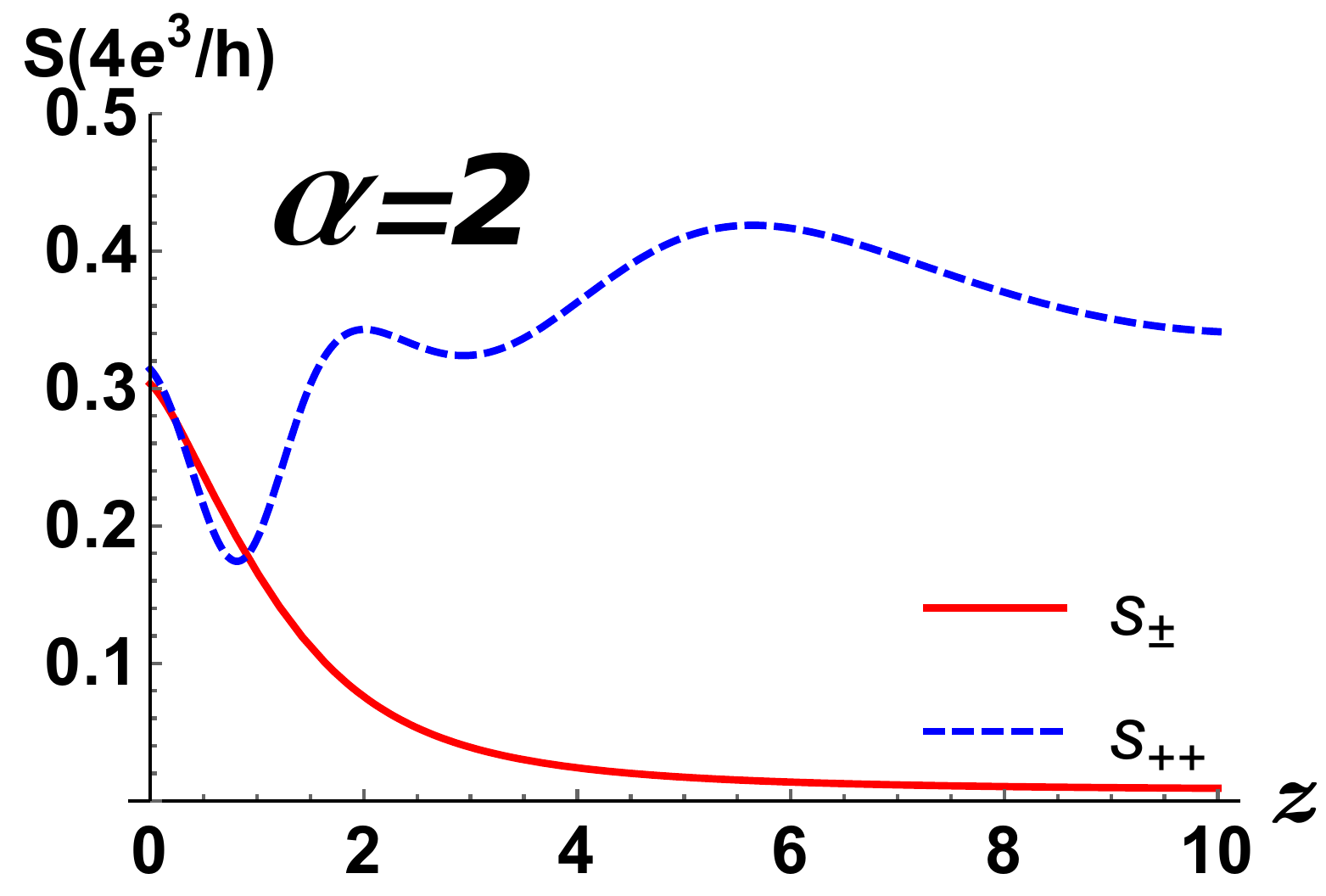} }  
  \hspace{-0.5cm}      \subfigure[]
      { \includegraphics[scale=0.28]{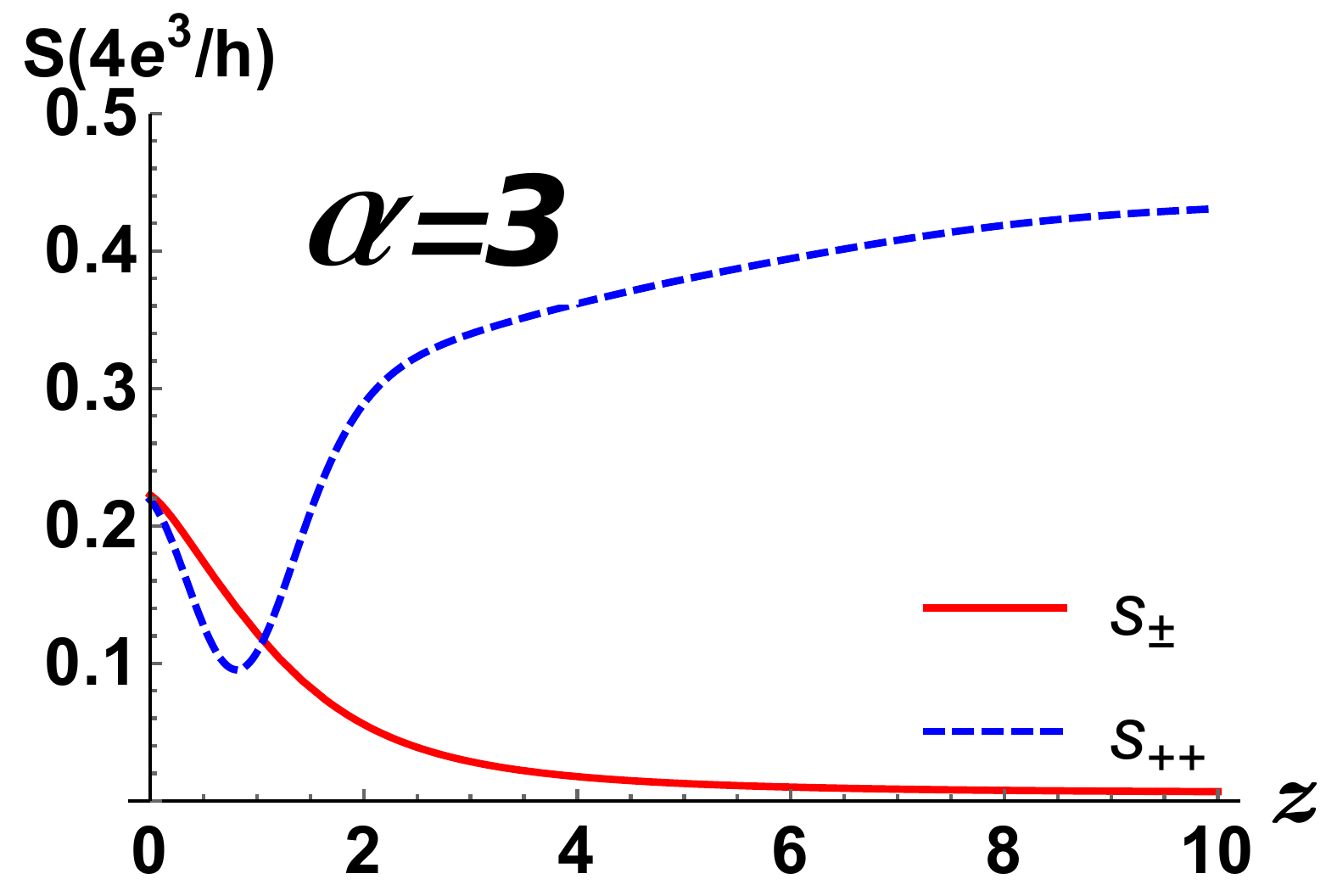} }
    \caption{Differential shot noise for a $N_1/I/N_2/I/Ip$ superconductor junction vs barrier strength $z$ with $\Delta_2$=3.75meV, $\Delta_1$=2.5meV, $E_F$=3.8eV, $a=10nm$, $z_1=z_2=z$ and $eV = \Delta_1 $ for (a) $\alpha=2$, (b) $\alpha=3$.}
\end{figure}
\begin{figure}[h]\hspace{-0.5cm}
        \subfigure[]
  {     \includegraphics[scale=0.275]{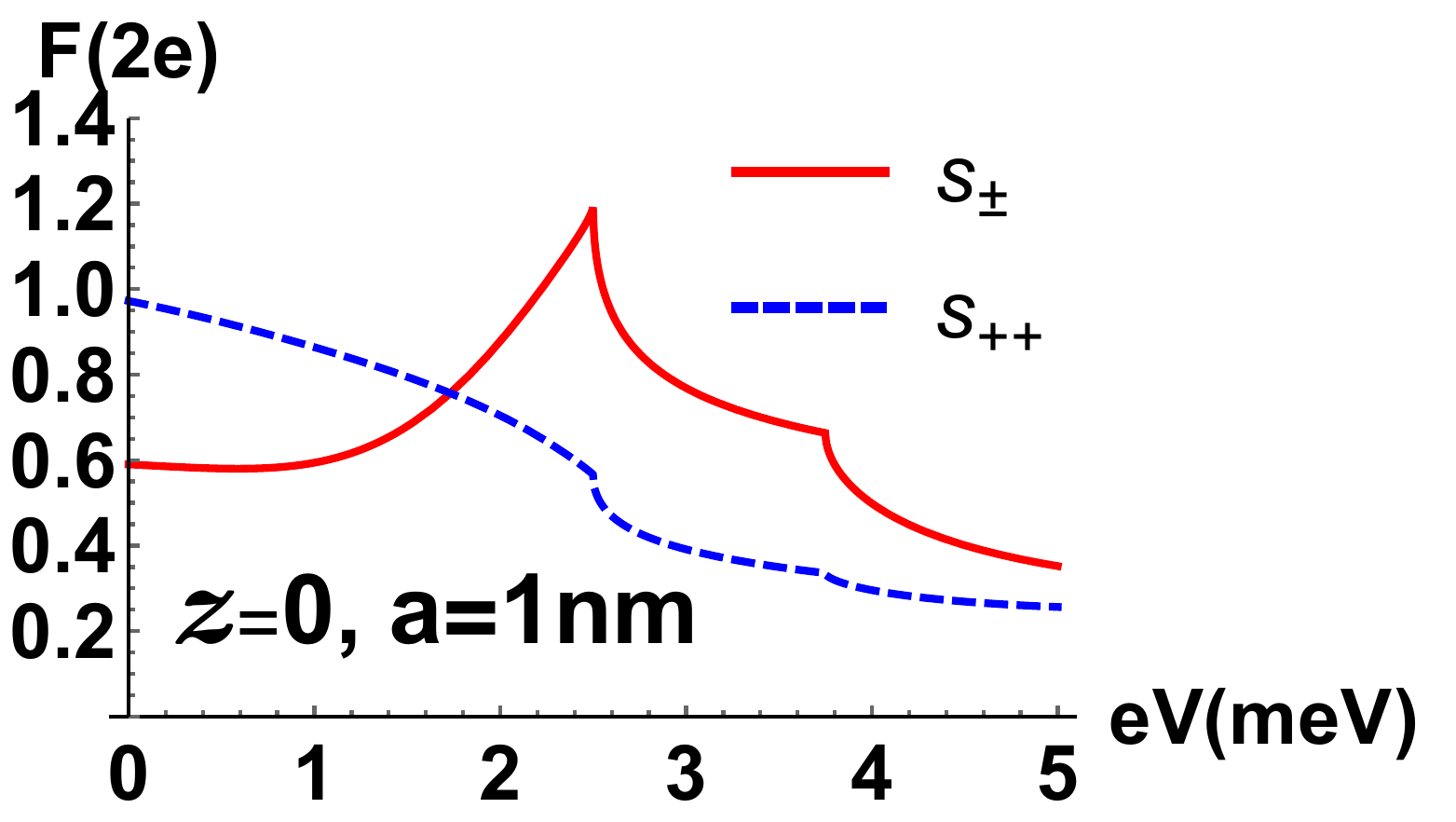} }
   \hspace{-0.5cm}     \subfigure[]
{   \includegraphics[scale=0.275]{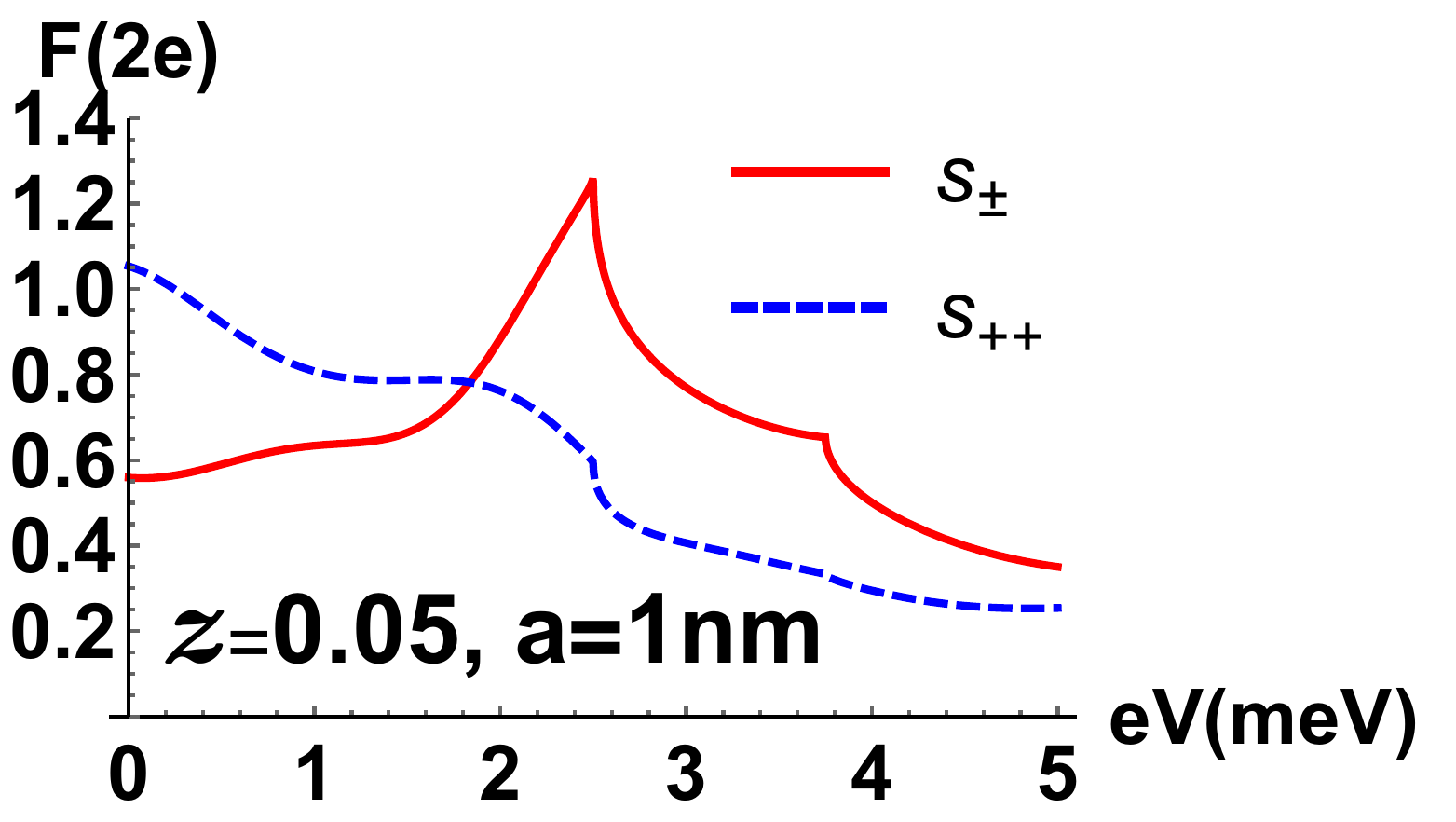} }
  \hspace{-0.5cm}      \subfigure[]  
  {    \includegraphics[scale=0.275]{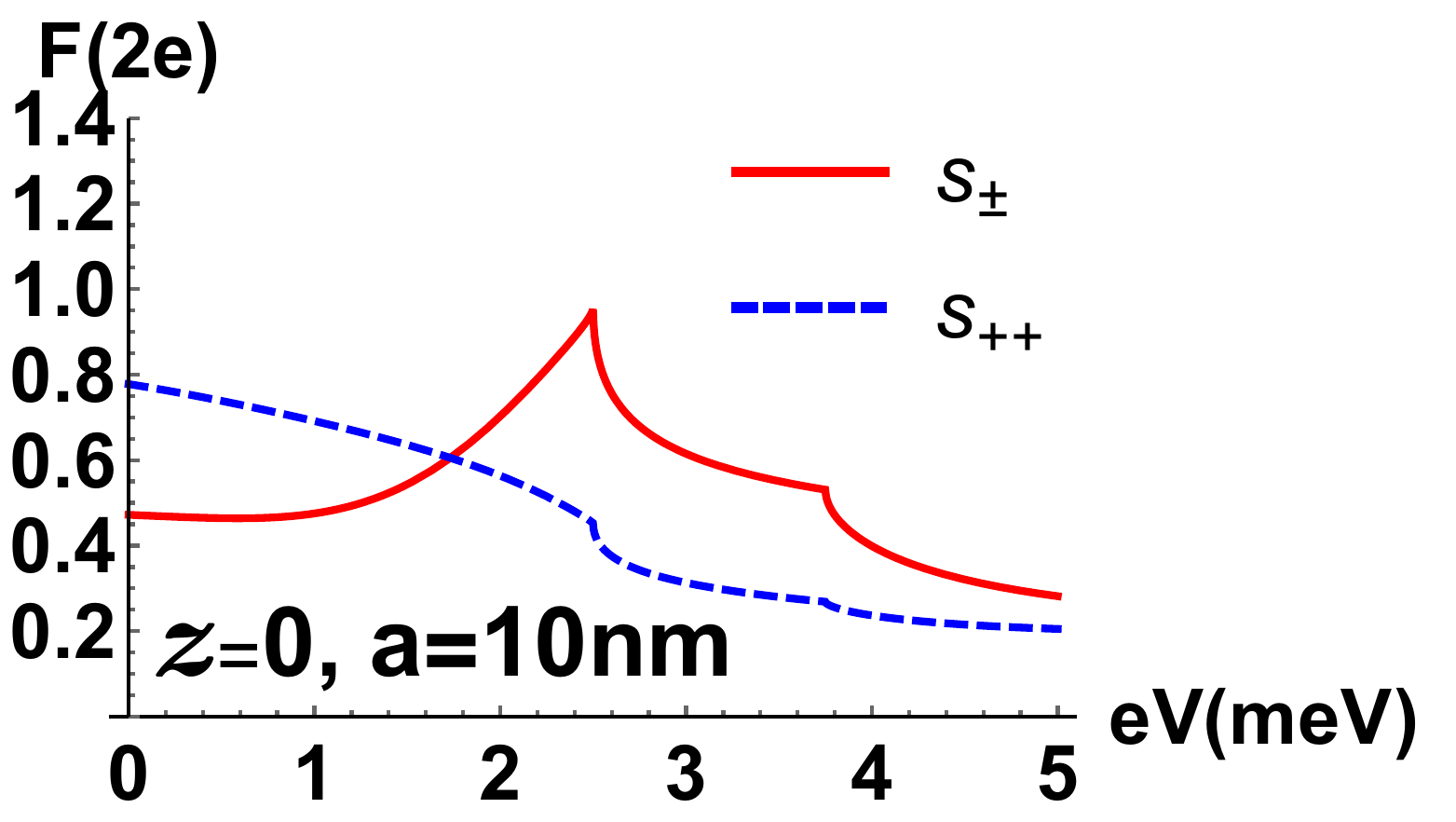}}
\caption{Differential Fano factor for a $N_1/I/N_2/I/Ip$ superconductor junction vs bias voltage V(meV) with $\Delta_2=3.75$meV, $\Delta_1=2.5$meV, $E_F=3.8$eV and $\alpha=1$ for (a) $a=1nm$ and $z=0$, (b) $a=1nm$ and $z=0.05$, (c) $a=10nm$ and $z=0$. }
\end{figure}
In Fig.~4, we plot the differential shot noise with respect to barrier strength for different values of interband coupling strength($\alpha$). The differential shot noise for s$_\pm$ pairing tends to zero but for s$_{++}$ pairing tends to a finite value in the tunnel limit($z \rightarrow large$) regardless of any change in interband coupling strength($\alpha$).

In Fig.~5 for a $N_1/I/N_2/I/Ip$ junction the differential Fano factor for s$_\pm$ pairing increases with bias voltage and tends to super Poissonian values near $eV \simeq \Delta_1$ regardless of any small changes to thickness ($a$) and barrier strength($z$) as shown in Fig.~5(b) and Fig.~5(c). Near $eV \simeq \Delta_1$, s$_\pm$ pairing shows a peak while s$_{++}$ pairing symmetry shows a dip in the Fano factor. These results for conductance in a normal metal bilayer in proximity to a Ip superconductors are in contrast to that of a single layer wherein no such difference can be seen between s$_{++}$ and s$_{\pm}$(see, supplementary material for further details). Next we deal with the $FM/I/NM/I/Ip$ junction.
\section{$FM/I/NM/I/Ip$ superconductor junction}
\subsection{Differential conductance}
\begin{figure}[h]
\hspace{-0.5cm}
        \subfigure[]
{ \includegraphics[scale=0.275]{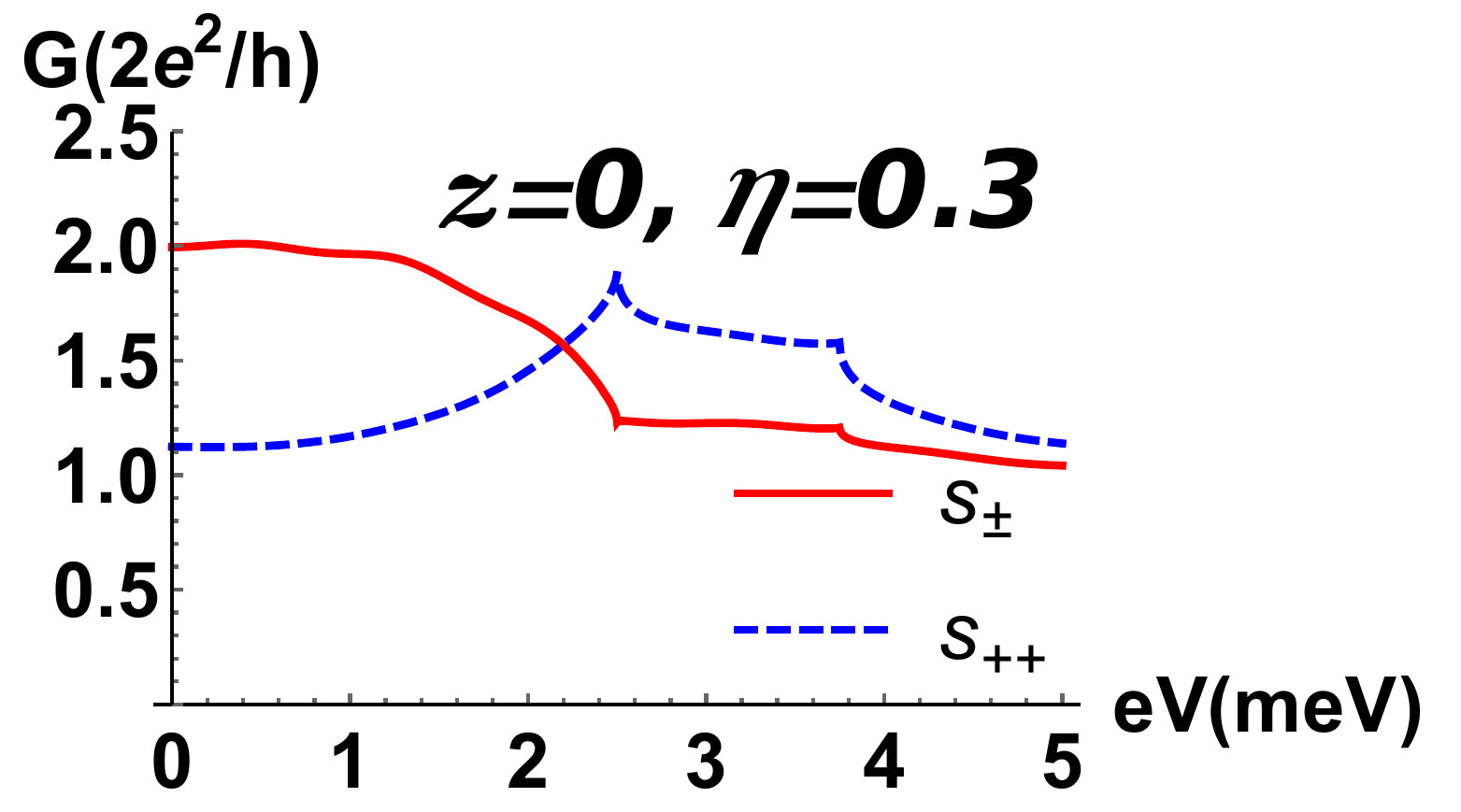} }   
   \hspace{-0.5cm}     \subfigure[]
{  \includegraphics[scale=0.275]{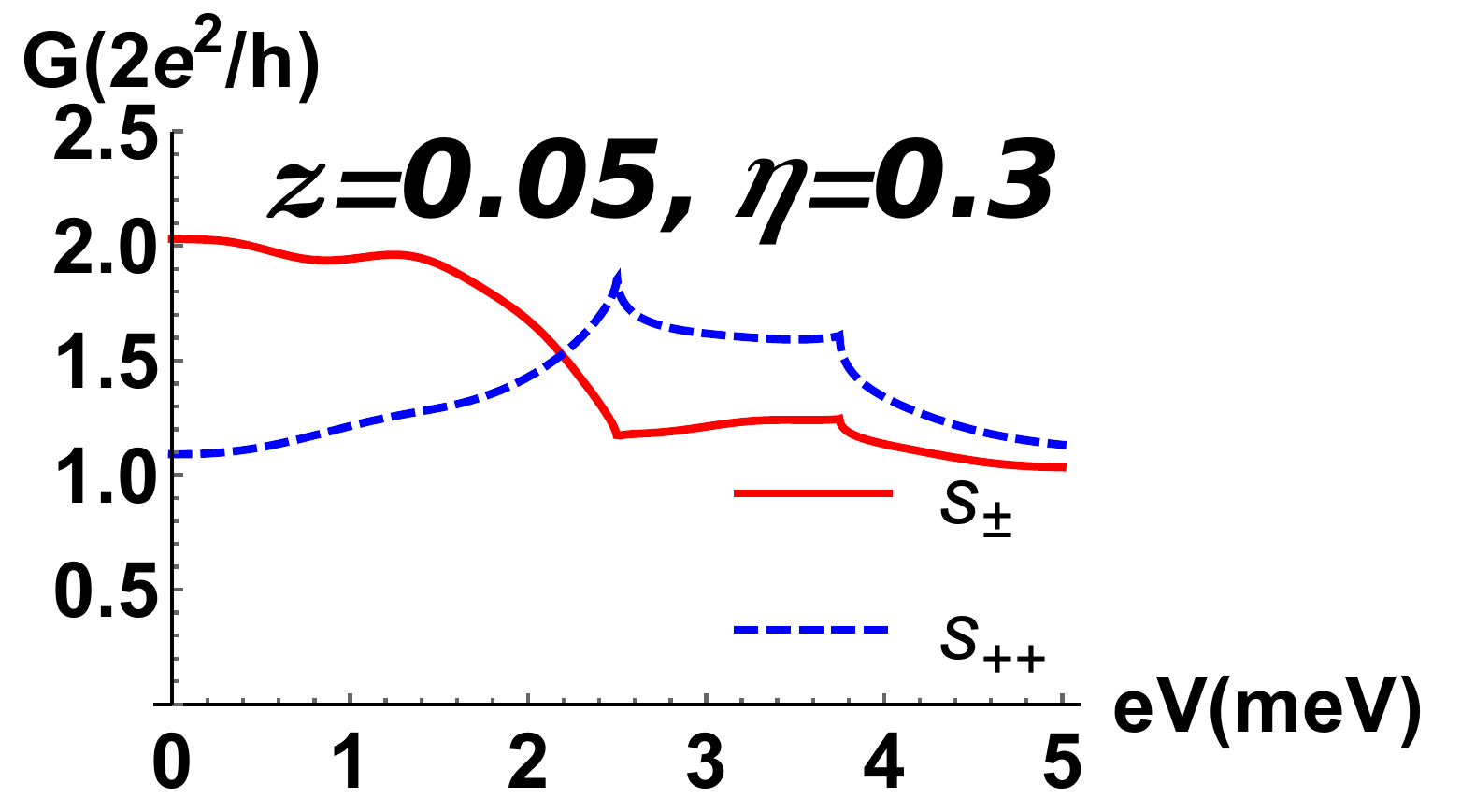} }
  \hspace{-0.5cm}      \subfigure[]
{  \includegraphics[scale=0.275]{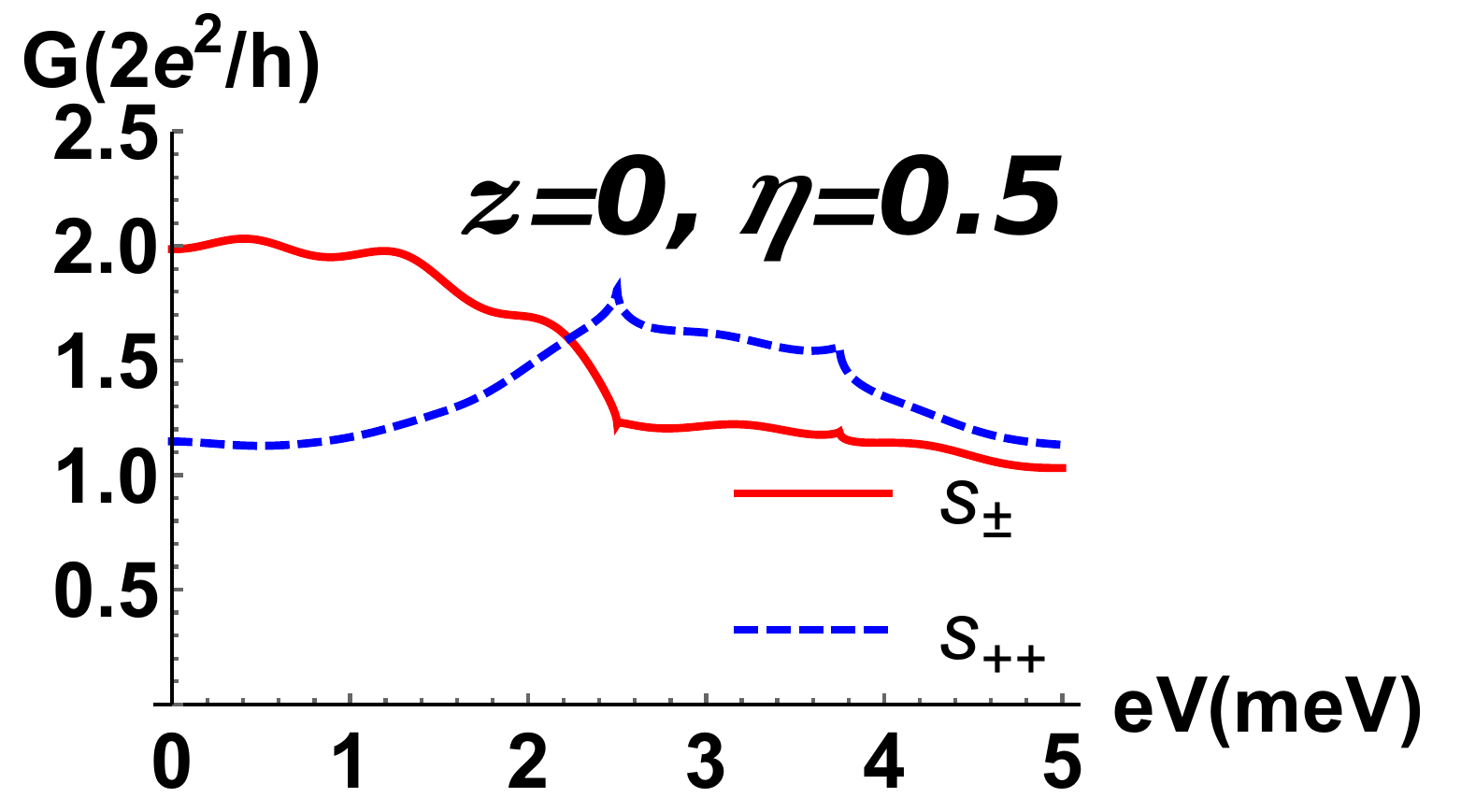}}
\caption{Normalized differential conductance for a $FM/I/NM/I/Ip$ superconductor junction vs bias voltage V(meV)  where $\Delta_2$=3.75meV, $\Delta_1$=2.5meV, $E_F$=3.8eV, $a=1$nm and $\alpha=1$ for (a) $\eta=0.3$ and $z=0$, (b) $\eta=0.3$ and $z=0.05$, (c) $\eta=0.5$ and $z=0$.}
\end{figure}
\begin{figure}[h]
\hspace{-0.5cm}
        \subfigure[]
{ \includegraphics[scale=0.275]{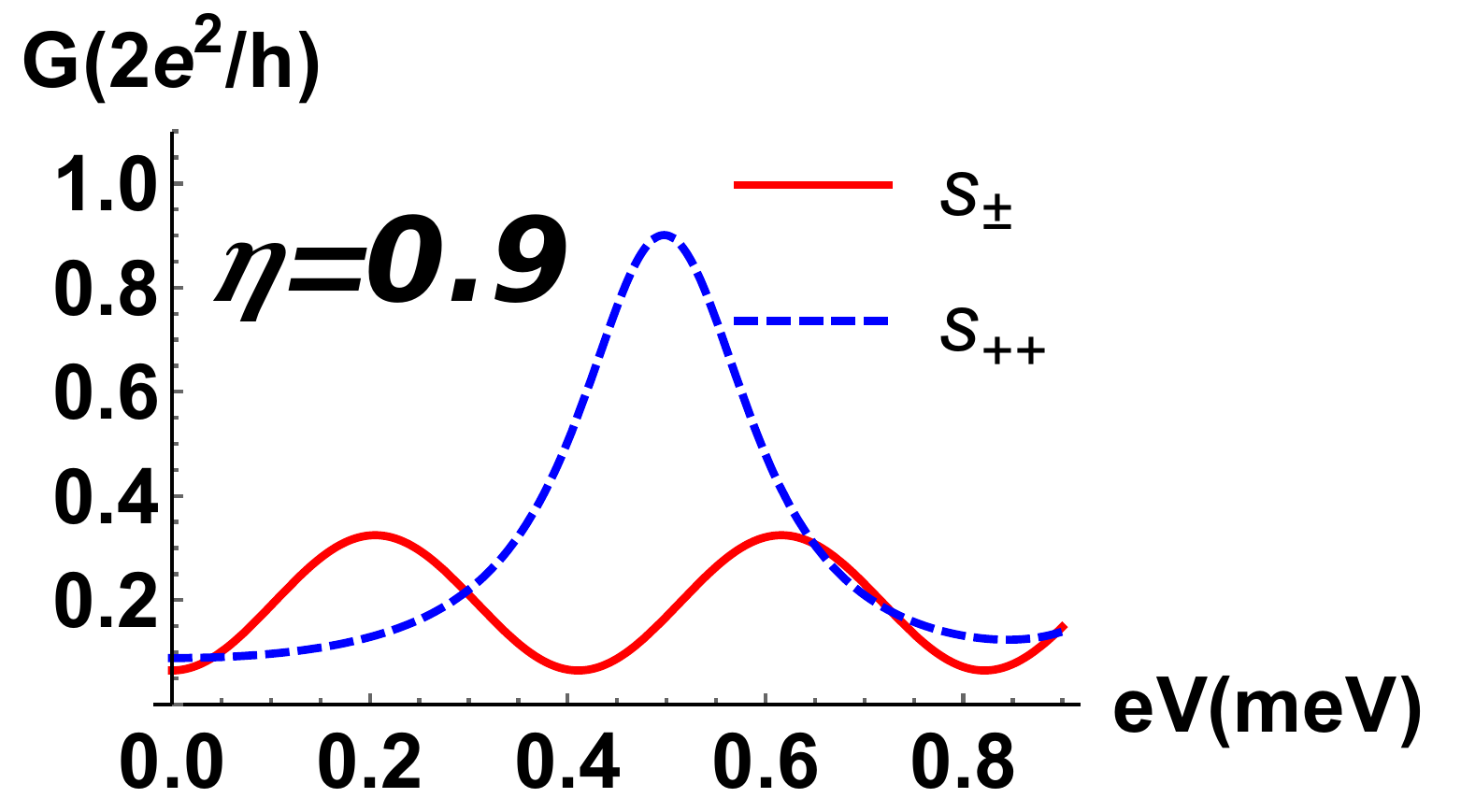} }
  \hspace{-0.5cm}      \subfigure[]
{ \includegraphics[scale=0.275]{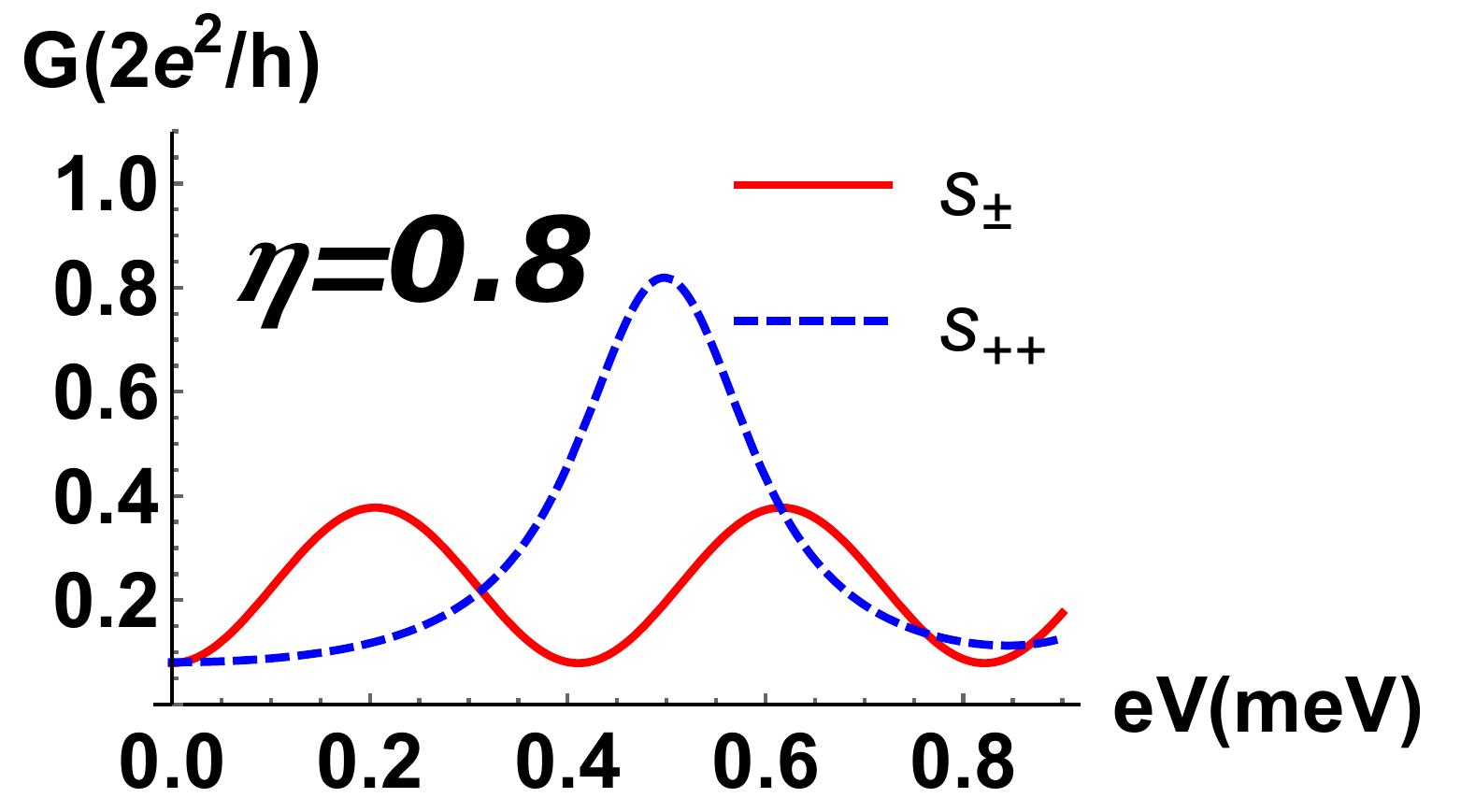} }    
 \hspace{-0.5cm}       \subfigure[]
{ \includegraphics[scale=0.275]{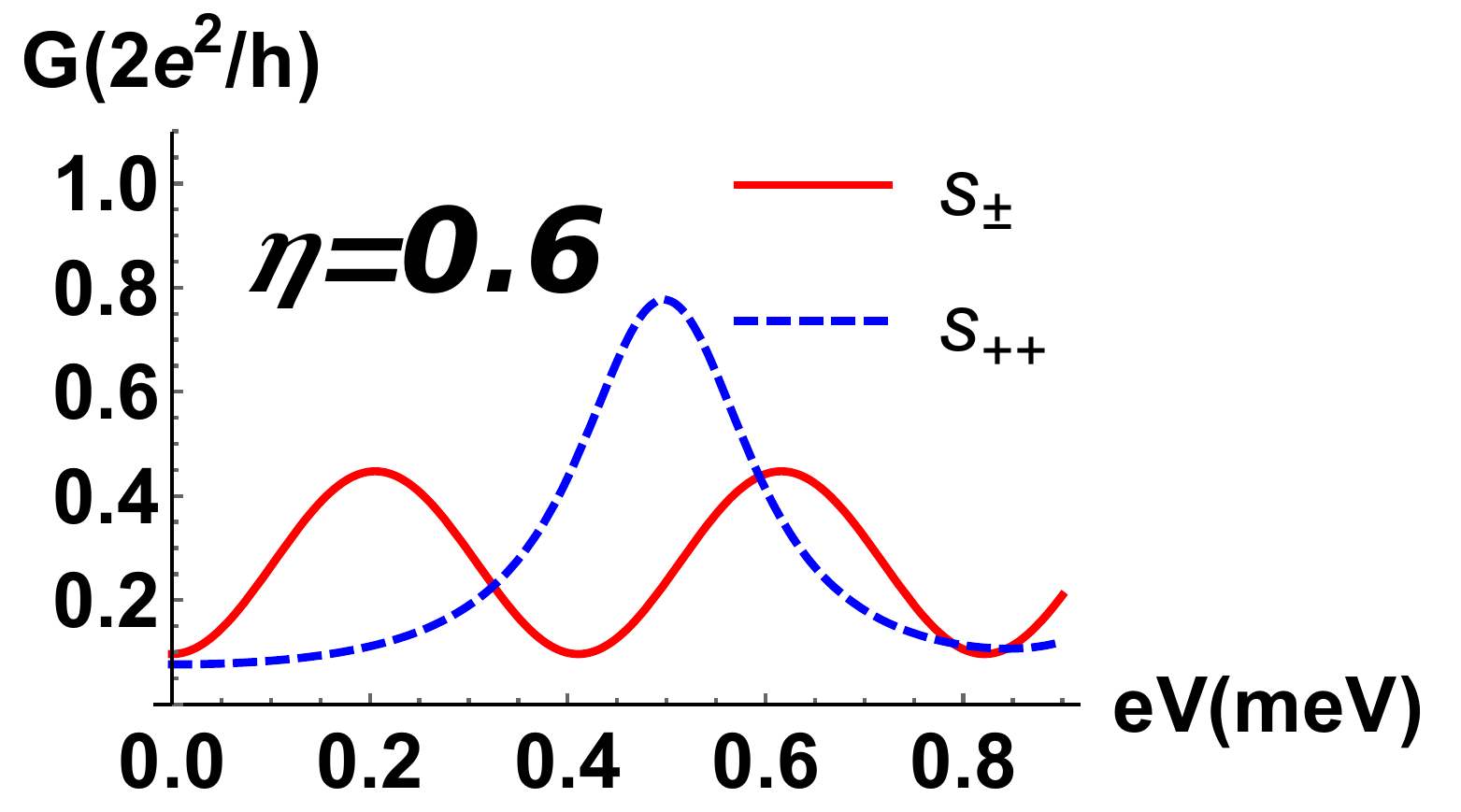}}
\caption{Normalized differential conductance for a $FM/I/NM/I/Ip$ superconductor junction vs bias voltage V(meV) with $\Delta_2$=3.75meV, $\Delta_1$=2.5meV, $E_F$=3.8eV, $a=1nm$, $z_2$=$z_1$=2 and $\alpha=$2 for (a) $\eta=0.9$, (b) $\eta=0.8$, (c) $\eta=0.6$.}
\end{figure}
In Fig.~6(a) differential conductance for a $FM/I/NM/I/Ip$ junction with s$_\pm$ pairing shows ZBCP while s$_{++}$ pairing shows a dip at zero bias for thickness($1nm$), interband coupling strength($\alpha=1.0$) and $\eta$=$h_0/E_F$=0.3(with magnetization $h_0$). A small change in barrier strength and magnetization does not affect the ZBCP in s$_\pm$ pairing. further, for s$_\pm$ pairing one sees a conductance peak at $eV= \Delta_1$ while s$_{++}$ pairing shows a dip as shown in Figs.~6(b,c).

In Fig.~8 we plot differential conductance vs bias voltage for $FM/I/NM/I/Ip$ superconductor junction for different magnetization values. The period of differential conductance oscillation for s$_{++}$ pairing is half of the period of conductance oscillation for s$_{\pm}$ pairing and this is irrespective of any change in magnetization in Ferromagnet. In Appendix of supplementary material we show that unlike a $FM/I/NM/I/Ip$ superconductor junction, there is no differential conductance oscillation for a $FM/I/Ip$ junction neither for s$_{++}$ pairing nor for s$_{\pm}$ pairing, this shows the advantage a bilayer in conjunction with Ip superconductor has over a single layer in determining the pairing symmetry.
\subsection{Differential shot noise}
\begin{figure}[h]
\hspace{-0.85cm}
        \subfigure[]
{ \includegraphics[scale=0.275]{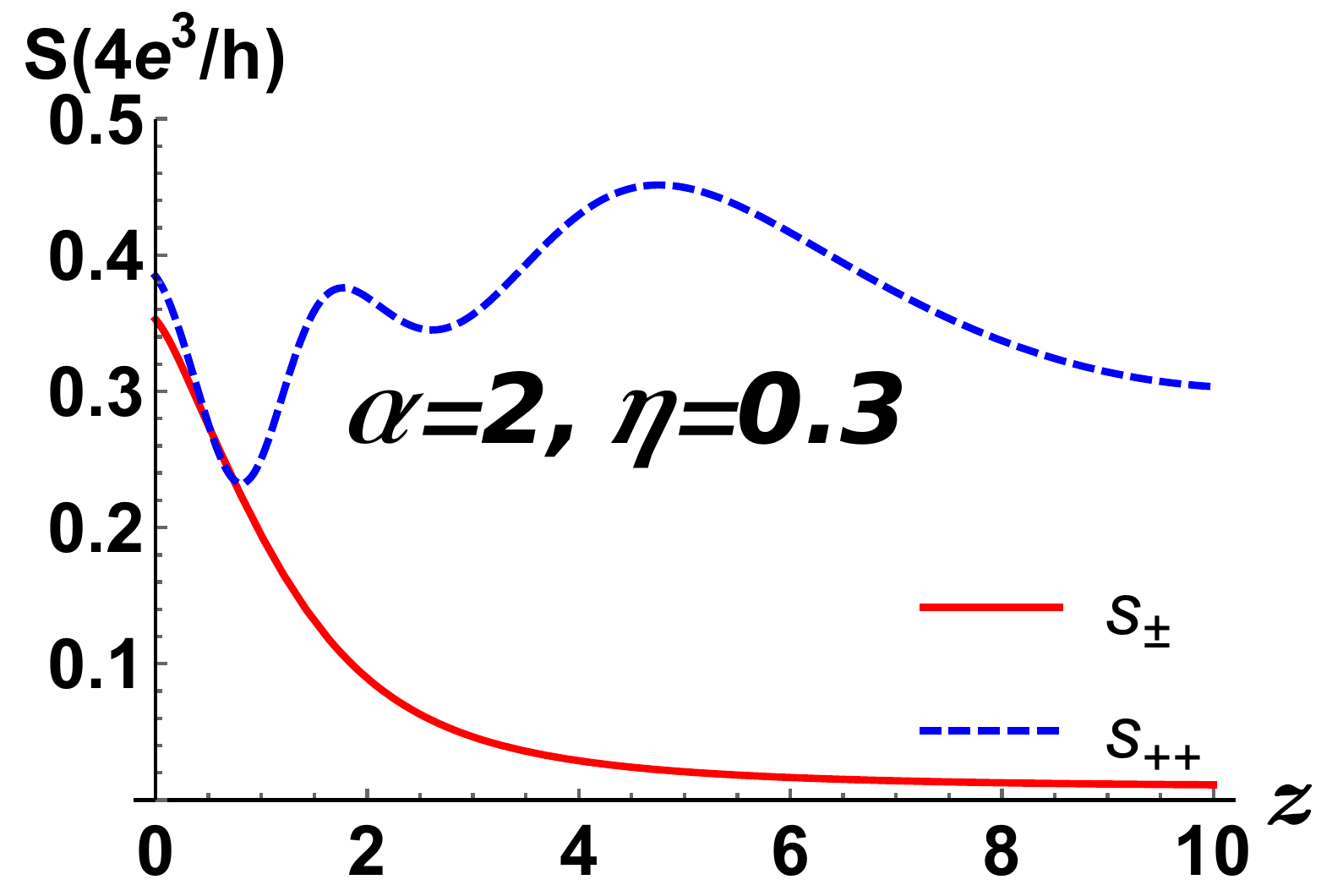} }       
    \hspace{-0.5cm}    \subfigure[]
{   \includegraphics[scale=0.275]{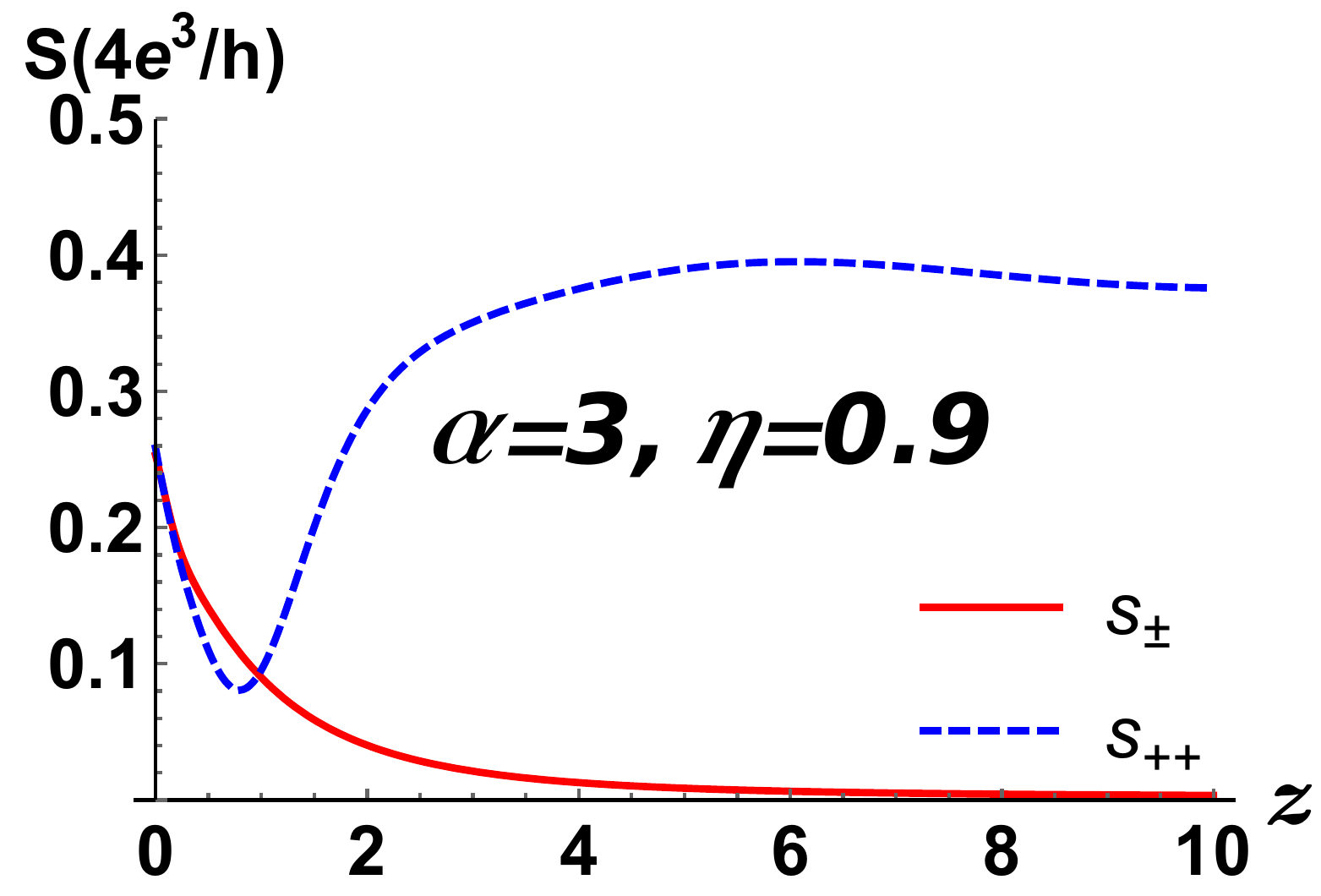} }
     \hspace{-0.5cm}   \subfigure[]
{     \includegraphics[scale=0.275]{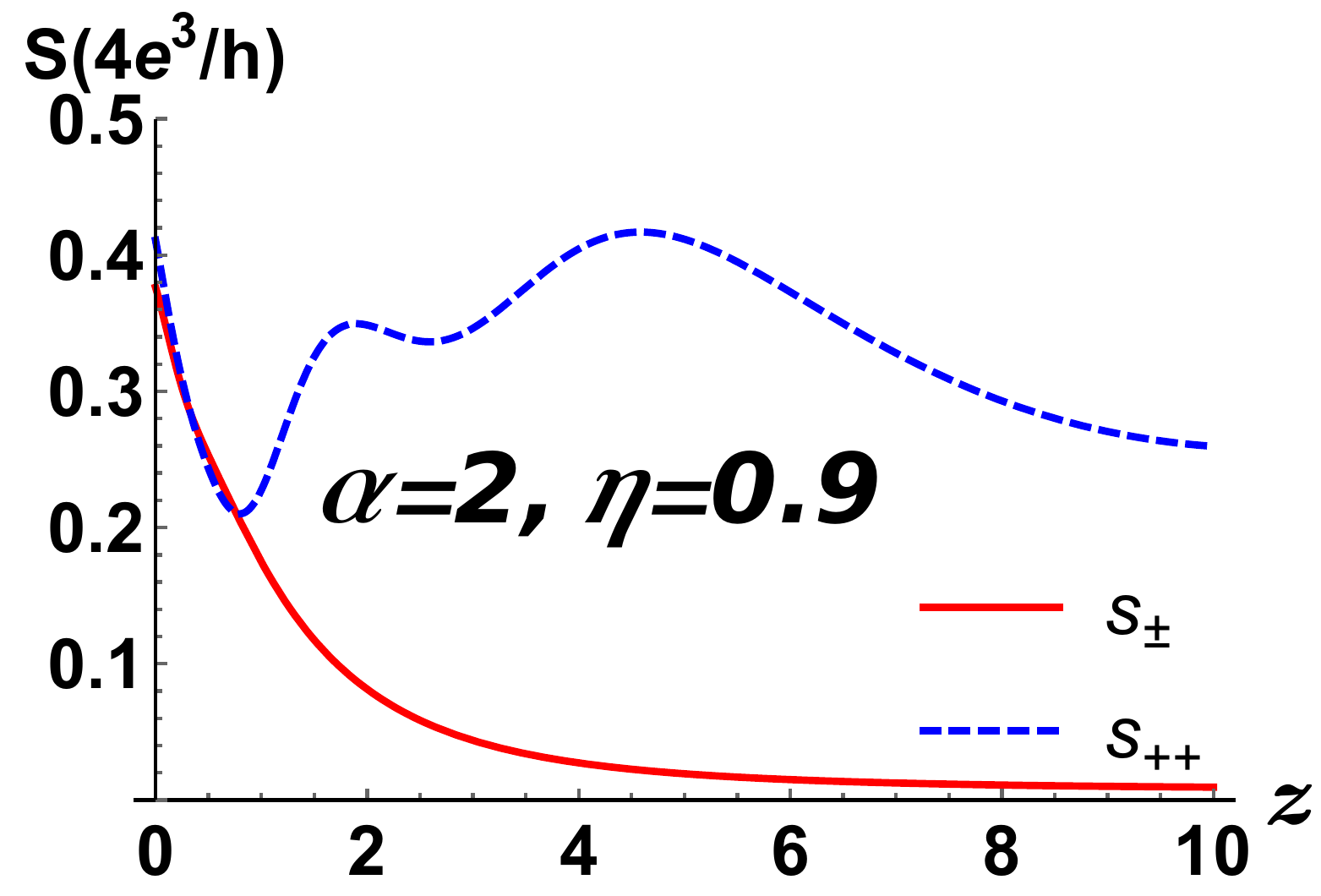}}
\caption{Differential shot noise for a $FM/I/NM/I/Ip$ superconductor junction  vs barrier strength $z$ with $\Delta_2$=3.75meV, $\Delta_1$=2.5meV, $E_F$=3.8eV, $a=10$nm, $z_1=z_2=z$ and $eV = \Delta_1$ for (a) $\alpha=2$ and $\eta=0.3$, (b) $\alpha=3$ and $\eta=0.9$, (c) $\alpha=2$ and $\eta=0.9$.}
\end{figure}
In Fig.~8 we plot differential shot noise for a $FM/I/NM/I/Ip$ junction for different values of interband coupling strength and magnetization in Ferromagnet. Differential shot noise in tunnel limit vanishes for s$_\pm$ pairing but tends to finite values for s$_{++}$ pairing regardless of any change in interband coupling strength and magnetization.
\section{ Experimental realization}
We have used shot noise as a probe to detect pairing symmetry of Ip superconductor while most of the other methods use the charge conductance or a Josephson junction. The shot noise gives us another tool to probe pairing symmetry and also to confirm the results from conductance and Josephson junctions. Further, in our setup of a bilayer normal metal in conjunction with Ip superconductor gives a zero bias conductance peak which is not seen in single normal metal layer in conjunction with Ip superconductor. Finally, our results for shot noise are valid in the tunnel limit which is an added advantage as we explain below. 

For Iron based superconductors, the surface quality strongly depends on the material as well as on the preparation method for the Josephson junction to be realized. Detection of the symmetry of the order parameter is not clear in most types of Josephson junctions like corner junction, hybrid junction, etc., see Ref.~\cite{seidel} for more details. Further, problems like surface roughness, chemical stability, etc., need to be overcome in order to detect the pairing symmetry of Ip superconductor using Josephson junctions as explained in Ref.~\cite{seidel}. Finally, to experimentally detect pairing symmetry using Josephson junctions one needs a thin insulating layer between the two Ip superconductors, which is difficult to fabricate precisely, see Ref.~\cite{seidel}. However, in our paper we needn't fabricate precisely a thin interface insulating layer between normal metal and Ip superconducting layer as in our proposal it is in the tunnel limit that the difference in shot noise between $s_{++}$ and $s_{\pm}$ pairing are stark, so this is no longer a problem.

Thus, while thin barriers are difficult to fabricate due to high precision requirements. Thick barriers, i.e., the tunnel limit are relatively easier to design. One way to do this is via oxidation. Higher oxygen pressure with longer oxidation time results in a thicker oxide barrier, see Ref.~\cite{wang}. Another parameter that is tuned in our proposal is the bias voltage($E/\Delta_{1}$). Experimentally, bias voltage can be tuned to check the pairing symmetries of Ip superconductor from zero to any arbitrary bias. In figures of this paper we have the upper limit at  $2 \Delta_{1}$,(where $\Delta_{1}=$ superconducting gap for band 1) and in our paper we take $\Delta_{2} / \Delta_{1} = 1.5$. Further, doping  Ip superconductors can tune the inter-band coupling strength ($\alpha$) as mentioned in Ref.~\cite{liu}. Experimentally, doping  Cobalt atoms in the Ip superconductor Ba(Fe$_{1-x}$Co$_{x})_2$As$_{2}$ can tune the interband coupling to strong coupling at optimal doping, see Ref.~\cite{hardy}. This is important as in our paper we deal with both strong ($\alpha=3$) as well as weak($\alpha=2$) inter-band coupling, see Fig.~4 of our manuscript. 
\section{Conclusion}
Part of the difficulty in determining pairing symmetry of Iron-based superconductors is that different experiments seem to shows different results in different doping regimes and in different compounds\cite{RPP}. In certain samples, a small non zero resistance has been observed below $T_c$ due to the presence of inter-growth defect\cite{nature} that may affect the experimental results in Josephson junctions. {Having said this the Andreev conductance measured in SNS contacts, see Ref.~\cite{kuznech}, shows that as bias voltage decreases the differential conductance falls sharply and then increases giving a peak near zero bias. We too in our work see that s$_{\pm}$ pairing symmetry shows peak at zero bias voltage and decreases with increase of the bias voltage. This shows that  conclusions of our work can be experimentally measured both the differential conductance as well as shot noise. However, no experimentalist has done the shot noise measurements, our work hopefully will motivate experimentalists to look at signatures of the pairing symmetry of Ip superconductors via shot noise measurements.} Real measurements are often influenced by thermal noise, which smears the shape of the current near the critical current\cite{PSS}. We can avoid these difficulties by calculating the shot noise in the tunnel limit, i.e., at $z \rightarrow$ large, where differential shot noise vanishes for s$_{\pm}$ pairing but is finite for s$_{++}$ pairing with thickness of intermediate layer($10nm$) regardless of any change in interband coupling strength and magnetization as shown in Table I and table II. 
\begin{table}[!]
\centering
 \caption{$s_{++}$ vs. $ s_\pm$ pairing for $N_1/I/N_2/I/Ip$ junction.}
\resizebox{\columnwidth}{!}{
\begin{tabular}{|c|c|c|c|}
\hline
\begin{tabular}[c]{@{}c@{}}Pairing \\ symmetry\end{tabular} & \begin{tabular}[c]{@{}c@{}}Differential conductance-\\ Zero biased conductance \\ peak \end{tabular} & \begin{tabular}[c]{@{}c@{}}Differential shot noise\\   (Tunnel limit)\end{tabular} & \begin{tabular}[c]{@{}c@{}}Differential \\ Fano factor\end{tabular}                      \\ \hline
$\mathbf{s_{++}}$                                                         & No                                                                                     & Tends to finite value                                                                          & \begin{tabular}[c]{@{}c@{}}Sub-Poisonian \\ near $eV= \Delta_1$ \end{tabular}   \\ \hline
$\mathbf{s_\pm}$                                                         & Yes                                                                                    & Vanishes                                                                         & \begin{tabular}[c]{@{}c@{}}Super-Poisonian \\ near $eV= \Delta_1$ \end{tabular} \\ \hline
\end{tabular}}
\end{table}
\begin{table}[!]
\centering
 \caption{$s_{++}$ vs. $ s_\pm$ pairing for $FM/I/N/I/Ip$ junction}
\resizebox{\columnwidth}{!}{\begin{tabular}{|c|c|c|c|}
\hline
\begin{tabular}[c]{@{}c@{}}Pairing \\ symmetry\end{tabular} & \begin{tabular}[c]{@{}c@{}}Differential conductance-\\ Zero biased conductance \\ peak \end{tabular} & \begin{tabular}[c]{@{}c@{}}Differential conductance\\   (Period of oscillation)\end{tabular} & \begin{tabular}[c]{@{}c@{}}Differential shot noise\\   (Tunnel limit)\end{tabular} \\ \hline
$\mathbf{s_{++}} $                                                        & No                                                                                     & \begin{tabular}[c]{@{}c@{}}Half the period of\\  oscillation of $s_\pm$ \end{tabular}                 & \begin{tabular}[c]{@{}c@{}}Tends to finite value \end{tabular}   \\ \hline
$\mathbf{s_\pm}$                                                         & Yes                                                                                    & \begin{tabular}[c]{@{}c@{}}Twice the period of\\  oscillation of $s_{++}$ \end{tabular}             & \begin{tabular}[c]{@{}c@{}}Vanishes \end{tabular} \\ \hline
\end{tabular}}
\end{table}
In Table I and II we have summarized the main results to distinguish between s$_{++}$ and s$_{\pm}$ pairing, as already shown in the plots for both $N_1/I/N_2/I/Ip$ and $FM/I/N/I/IP$ junction. In Table I, for a $N_1/I/N_2/I/Ip$ junction s$_{\pm}$ pairing shows ZBCP unlike s$_{++}$ pairing. The differential shot noise vanishes in the tunnel limit for s$_{\pm}$ pairing while tends to finite value for s$_{++}$ pairing. the differential Fano factor for s$_{++}$ pairing tends to sub-Poisonian value while for s$_{\pm}$ pairing tends to super-Poisonian value near $eV = \Delta_1$. In Table II, for a $FM/I/NM/I/Ip$ junction s$_{\pm}$ pairing shows ZBCP unlike s$_{++}$ pairing. The period of conductance oscillation for s$_{\pm}$ pairing is half the period of conductance oscillation of s$_{++}$ pairing which is the unique silver bullet to probe the pairing symmetry. Regardless of whether we use normal metal or ferromagnet, the differential shot noise in the tunnel limit vanishes for s$_{\pm}$ pairing, while it is finite value for s$_{++}$ pairing.

{\bf Acknowledgments:} C.B. wishes to thank DAAD, Germany for a research stay at Aachen in summer 2016 where this project got underway. We acknowledge Fabian Hassler for his crucial insights at the initial stage of the project. This work was supported by the grants- 1. Josephson junctions with strained Dirac materials and their application in quantum information processing from SERB, New Delhi,
Government of India, Grant No. CRG/20l9/006258, and 2. Nash equilibrium versus Pareto optimality in N-Player games, SERB MATRICS Grant No. MTR/2018/000070.

\onecolumn
\section{Supplementary Material}
In this accompanying supplementary material, we first deal with the wave functions and boundary conditions for the Metal-Insulator-Metal-Insulator-Iron  pnictide superconductor junction $N_{1}/I/N_{2}/I/Ip$ as well as Ferromagnet-Insulator-Metal-Insulator-Iron  pnictide superconductor junction $FM/I/NM/I/Ip$ junction.  Finally, we study the main advantages between a  bi-layer normal metal or ferromagnet-normal metal bi-layer in conjunction with Iron Pnictide superconductor over a single layer normal metal or single feromagnetic layer in conjunction with Iron Pnictide superconductor. 
 
\subsection{Wavefunctions and Boundary Conditions for $N_{1}/I/N_{2}/I/Ip$ superconductor junction} 
The wavefunctions in metal $N_1$ and $N_2$ are $\psi_{N_1}$ and $\psi_{N_2}$. The $N_1/I/N_2/I/Ip$ superconductor junction has insulators at $x=-a$ and $x=0$, the two insulators are described by $\delta$-function potentials: $U(x)=U_1 \delta(x+a) + U_2 \delta(x)$ with $U_1$ and $U_2$ being the barrier strengths. The Iron based superconductor possesses two superconducting gaps $\Delta_{1,2}$ in both the bands $\Gamma$ and $M$\cite{RPP}. The superconducting phases of the gaps are $\phi_1$ and $\phi_2$. The $s_{\pm}$ pairing model has unequal gaps ($\Delta_1 \ne \Delta_2$) with phases of opposite signs, i.e., $\phi_1 - \phi_2 = \pi$, while $s_{++}$ pairing model has unequal gaps ($\Delta_1 \ne \Delta_2$) but with same sign, i.e., $\phi_1 = \phi_2$.

Similar to the Iron pnictide junction, we consider the metals $N_1$ and $N_2$ to have two distinct bands with the band energies as was also done in Ref.~\cite{linder}, $\varepsilon_{k,1}$ = $(\hbar^2/2m)(k_F-\pi)^2-E_F$ and $\varepsilon_{k,2}$ = $(\hbar^2/2m)(k_F-\pi)^2+E_F$ as in Fig.~1 {\em of main manuscript}. Further, we assume the hole and electron Fermi surfaces to be circular and of same size although in actuality they aren't exactly circular. We have relaxed the condition of Andreev approximation, however the additional phase shift in the first band makes no difference to the results at all and therefore in the subsequent calculation we neglect this additional phase shift.

From Eq.~(3) {\em of main manuscript}, the wave functions in the three regions when an electron is incident from the left in band $1$ is-
{\hspace{-0.5cm}\small
\begin{eqnarray}
\psi_{N_1}(x)\!\!\!&=&\!\!\!\!\left(\begin{array}{c}
1\\
0\\
0\\
0
\end{array}\!\!\right)(e^{i k^+_{NM} x}\!+\!b_1 e^{-i k^+_{NM} x})\!+\!a_{1}\left(\begin{array}{c}
0\\
1\\
0\\
0
\end{array} \right) e^{i k^-_{NM} x}\!+\!b_{2}
\left(\begin{array}{c}
0\\
0\\
1\\
0
\end{array} \!\!\right) e^{-i k^+_{NM} x}\!+\!a_{2}\left(\begin{array}{c}
0\\
0\\
0\\
1
\end{array} \!\!\right) e^{i k^-_{NM} x},\mbox{for $x < -a$},\\
\psi_{N_2}(x)\!\!&=&\!\!\left(\begin{array}{c}
1\\
0\\
0\\
0
\end{array}\right) (t_1 e^{i k^+_{NM} x} + g_1 e^{-i k^+_{NM} x}) + 
\left( \begin{array}{c}
0\\
1\\
0\\
0
\end{array} \right) (h_1 e^{i k^-_{NM} x} + f_1 e^{-i k^-_{NM} x}) +
\left( \begin{array}{c}
0\\
0\\
1\\
0
\end{array} \right) (t_2 e^{i k^+_{NM} x} + g_2 e^{-i k^+_{NM} x}) \nonumber  \\
&+& 
\left(\begin{array}{c}
0\\
0\\
0\\
1
\end{array} \right) (h_2 e^{i k^-_{NM} x} + f_2 e^{-i k^-_{NM} x}), \mbox{or $-a < x < 0$, and} \\
\psi_{IP}(x)\!\!\!&=&\!\!\!c_1\!\!\left(\!\!\begin{array}{c}
u_1\\
v_1 e^{-i \phi_1}\\
0\\
0
\!\!\end{array}\right)\!\! e^{i k^+_{1} x}\!+\!d_{1}\!\!\left(\!\!\begin{array}{c}
v_1 e^{-i \phi_1}\\
u_1\\
0\\
0
\end{array}\!\!\right)\!\!e^{-i k^-_{1} x}\!+\!c_{2}\!\!\left(\!\! \begin{array}{c}
0\\
0\\
u_2\\
v_2 e^{-i \phi_1}\end{array}\!\!\right)\!\!e^{i k^+_{2} x}\!+\!d_{2}\!\!\left(\!\!\begin{array}{c}
0\\
0\\
v_2 e^{-i \phi_1}\\
u_2
\end{array}\!\!\right)\!\!e^{-i k^-_{2} x}, \mbox{for $x > 0$}.
\end{eqnarray}
}
In Eqs.(1-3) above, the coherence factor for the gaps $\Delta_{1(2)}$ are $u_{1(2)}= \sqrt[]{(1/2)(1+\Omega_{1(2)}/E)}$, $v_{1(2)}= \sqrt[]{(1/2)(1-\Omega_{1(2)}/E)} $, with $\Omega_{1(2)} = \sqrt[]{E^2-\Delta_{1(2)}^2}$. $k^{\pm}_{NM}$ is the wave vector of an electron(hole) in the normal metal region defined as $k^{\pm}_{NM} \simeq k_F (1 \pm E/2 E_F)$ where the Fermi wave vector is $k_F = \frac{\sqrt[]{2 m E_F}}{\hbar}$ and the electron/hole energy level $E$. $k^{\pm}_{1(2)}$ is the wave vector of an electron(hole) like quasiparticle in Iron pnictide region defined as $k^{\pm}_{1(2)}\simeq k_F (1 \pm \Omega_{1(2)}/2 E_F)$. The Fermi energy and superconducting gaps are taken as $E_F=3.8$eV, $\Delta_{2}=3.75$meV and $\Delta_{1}=2.5$meV; see also~\cite{12}.

For an incoming electron from band $2$, the wave function of normal metal($N_1$) is simply obtained by letting [1,0,0,0] $e^{ik^+_{NM}x}$ go to [0,0,1,0] $e^{ik^+_{NM}x}$ in Eq.~(1). Here, $\{b_1, a_1 \}$ are the normal and Andreev reflection scattering amplitudes for band $1$, similarly we have $\{b_2, a_2 \}$-the normal and Andreev reflection scattering amplitudes for band $2$. The general boundary conditions at the interfaces can then be found from Fig.~2 of main manuscript as:
\begin{eqnarray}
\Psi_{N_1}|_{x=-a}& =& \Psi_{N_2}|_{x=-a},\\
\frac{\partial}{\partial x}(\Psi_{N_2} - \Psi_{N_1})|_{x=-a}&=& 2m\left( U_1 diag(\hat{1},\hat{1}) \right) \Psi_{N_1}|_{x=-a},\\
\Psi_{N_2}|_{x=0} &=& \Psi_{IP}|_{x=0},\\
\frac{\partial}{\partial x}(\Psi_{IP} - \Psi_{N_2})|_{x=0}&=& 2m\left( U_2 diag(\hat{1},\hat{1}) + \alpha_{0} \hspace{0.2cm} offdiag(\hat{1},\hat{1}) \right) \Psi_{N_2}|_{x=0},
\end{eqnarray}
using which all the scattering amplitudes can be determined. In Eqs.~(4-7) $\hat{1}$ is the $2\times2$ unit matrix and $diag$ and $offdiag$ denote diagonal and off-diagonal $4\times4$ matrices in which these unit matrices are embedded\cite{linder}. At this point we also introduce two dimensionless parameters characterizing the system, namely the barrier strength $z_i= 2m U_i/k_F$, $i = 1,2$ and the interband coupling strength $\alpha = 2m \alpha_{0} / k_F$. From the scattering amplitude $a_i,b_i$, $i=1,2$ we get the Andreev and normal reflection probabilities as $A_{\sigma}=|a_{\sigma}|^2$, $B_{\sigma}=|b_{\sigma}|^2$ where $\sigma =1,2$. This procedure of solving the boundary conditions (Eqs.~4-7) is repeated for an electron incident in band $2$ of metal $N_1$.
\section{Wave functions and boundary conditions in a Ferromagnet-Insulator-Metal-Insulator-Iron pnictide Superconductor junction}
\begin{figure}[h]
      \includegraphics[scale=0.5]{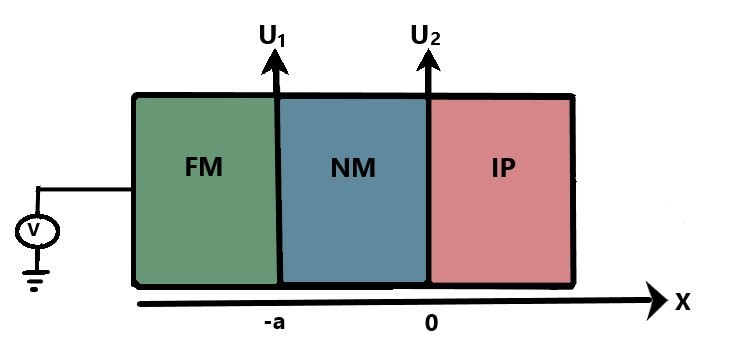}
      \caption{ Ferromagnet-Insulator-Normal metal-Insulator-Iron pnictide junction }
      \label{fig3}
   \end{figure}
The Ferromagnet-Insulator-Metal-Insulator-Iron pnictide ($FM/I/NM/I/Ip$) superconductor setting is shown in Fig.~\ref{fig3}, with wave functions: $\psi_{FM}(x)$, $\psi_{NM}(x)$ and $\psi_{IP}(x)$ for the ferromagnet, normal metal and Iron pnictide segments. For a spin up electron incident at the interface from left in band $1$, the resulting wavefunctions in various segments are: 
{\small
\begin{eqnarray}
\!\!\!\!\!\psi_{FM}(x)\!\!\!\!\!\!\!&=&\!\!\!\!\!\!\!\left(\!\!\begin{array}{c}
1\\
0\\
0\\
0
\end{array}\!\!\right)(e^{i k^+_{FM,\uparrow}x}\!\!+\!\!b_1 e^{-i k^+_{FM, \uparrow}x})\!\!+\!\!a_{1}\!\!\left(\!\!\begin{array}{c}
0\\
1\\
0\\
0
\end{array}\!\!\right)e^{i k^-_{FM,\downarrow}x}\!\!+\!\!b_{2}
\!\!\left(\!\!\begin{array}{c}
0\\
0\\
1\\
0
\end{array}\!\!\right)e^{-i k^+_{FM,\uparrow}x}\!\!+\!\!a_{2}\!\! \left(\!\!\begin{array}{c}
0\\
0\\
0\\
1
\end{array}\!\!\right) e^{i k^-_{FM, \downarrow} x},\mbox{for $x < -a$},\\
 \psi_{NM}(x)\!\!&=&\!\!\left(\begin{array}{c}
1\\
0\\
0\\
0
\end{array}\right) (t_1 e^{i k^+_{NM} x}\!\!+\!\! g_1 e^{-i k^+_{NM} x}) + 
\left( \begin{array}{c}
0\\
1\\
0\\
0
\end{array}\right)(h_1 e^{i k^-_{NM} x}\!\!+\!\!f_1 e^{-i k^-_{NM} x}) +
\left(\begin{array}{c}
0\\
0\\
1\\
0
\end{array}\right)(t_2 e^{i k^+_{NM} x}\!\!+\!\!g_2 e^{-i k^+_{NM} x}) \nonumber \\
&+& 
\left( \begin{array}{c}
0\\
0\\
0\\
1
\end{array} \right) (h_2 e^{i k^-_{NM} x}\!\!+\!\!f_2 e^{-i k^-_{NM} x}), \mbox{for $-a < x < 0$, and }\\
\psi_{IP}(x)\!\!\!\!\!\!\!&=&\!\!\!\!\!\! c_1\!\!\left(\!\!\!\!\begin{array}{c}
u_1\\
v_1 e^{-i \phi_1}\\
0\\
0
\end{array} \!\!\!\!\right)\!\! e^{i k^+_{1} x}\!\!\!+\!
d_{1}\!\!\left(\!\!\begin{array}{c}
v_1 e^{-i \phi_1}\\
u_1\\
0\\
0
\end{array} \!\!\right)\!\! e^{-i k^-_{1} x}\!\! +\!\!
c_{2}
\!\!\left(\!\!\begin{array}{c}
0\\
0\\
u_2\\
v_2 e^{-i \phi_1}
\end{array}\!\!\!\right)\!\! e^{i k^+_{2} x}\!\!+\!\!
d_{2} \!\!\left(\!\!\! \begin{array}{c}
0\\
0\\
v_2 e^{-i \phi_1}\\
u_2
\end{array} \!\!\!\!\right)\!\! e^{-i k^-_{2} x}, \mbox{for $x > 0$}.
\end{eqnarray}
}
Similar to Eqs.~(8-10), we can write wavefunction resulting from electron incident in band $2$ too. The possible reflection amplitudes are $b_1-$ normal reflection in band $1$,\hspace{0.05cm} $b_2-$ normal reflection in band $2$,\hspace{0.05cm} $a_1-$ Andreev reflection in band $1$,\hspace{0.05cm} $a_2-$ Andreev reflection in band $2$. 

The wave vector of an electron(hole) in the ferromagnet region\cite{EPJ} is $k^{\pm}_{FM,\sigma} \simeq k_F (1 \pm (E + \sigma h_0)/2 E_F )$. For electron and hole spin up $\sigma=1$ and spin down $\bar{\sigma}=-1$. In the $FM/I/NM/I/Ip$ superconductor junction of Fig.~\ref{fig3}, magnetization is defined as $h(x)=h_0 \Theta(x+a)$ , where $h_0$ is the magnetization and $\Theta$ is the Heaviside step function\cite{zcdong} and ${t_{\sigma},f_{\sigma},g_{\sigma},h_{\sigma}}$ are the transmission amplitudes in band $\sigma$, wherein $\sigma= 1,2$ and $u_i,v_i$ with  $i = 1,2$ are the usual coherence factors defined as before with superconducting gap $\Delta_i$.  The boundary conditions at the interfaces are:
\begin{eqnarray}
\Psi_{FM}(x=-a) = \Psi_{NM}(x=-a),\\
\frac{\partial}{\partial x}(\Psi_{FM} - \Psi_{NM})|_{x=-a}= 2m\left( U_1 diag(\hat{1},\hat{1})\right) \Psi_{NM}(x=-a),\\
\Psi_{NM}(x=0) = \Psi_{IP}(x=0),\\
\frac{\partial}{\partial x}(\Psi_{IP} - \Psi_{NM})|_{x=0}= 2m\left( U_2 diag(\hat{1},\hat{1}) + \alpha_{0} \hspace{0.2cm}        offdiag(\hat{1},\hat{1})\right) \Psi_{NM}(x=0),
\end{eqnarray}

Similar to that for $N_{1}/I/N_{2}/I/Ip$ superconductor junction, here also we consider the dimensionless parameters to characterize the the barrier strength $z_i= 2m U_i/k_F$, $i = 1,2$ and the interband coupling strength $\alpha = 2m \alpha_{0} / k_F$.
From Eqs.~(11-14), all the scattering amplitudes can be determined when spin up/down electron is incident in band $1$ or $2$. From the coefficients we can get the probabilities of Andreev reflection and normal reflection as $A_1 =(k^-_{FM \bar{\sigma}}/k^+_{FM \sigma})|a_1|^2$. $B_1 =|b_1|^2$ for electron incident from left in band $1$ and $A_2 =(k^-_{FM \bar{\sigma}}/k^+_{FM \sigma})|a_2|^2$, $B_2 =|b_2|^2$ for electron incident from band $2$.
\subsection{Advantages of a bilayer over single layer in conjunction with Iron Pnictide superconductor}
In this section, we compare the results for the differential conductance and differential shot noise for Normal metal/Insulator/Normal metal/Insulator/Iron pnictide  superconductor ($N_1/I/N_2/I/Ip$) junction with that for a Normal metal/Insulator/Iron Pnictide  superconductor ($N/I/Ip$) junction. The plots bring out quite clearly the necessity for the set-up with a normal metal bilayer in conjunction with the Iron pnictide superconductor to discriminate between the $s_{++}$ and $s_{\pm}$ pairing symmetries vis-a-vis the single normal metal layer in conjunction with the Iron pnictide superconductor. $\Delta_{i}, i=1,2$ is superconducting gap for band $i$ and in all figures of this Appendix we take the ratio $\beta=\Delta_2 / \Delta_1 = 1.5$. Then we compare our results of differential conductance oscillation for a $FM/I/NM/I/Ip$ superconductor junction with $FM/I/Ip$ superconductor junction.
\begin{figure}[h]
\hspace{-0.51cm}\subfigure[]
    {\includegraphics[scale=0.478]{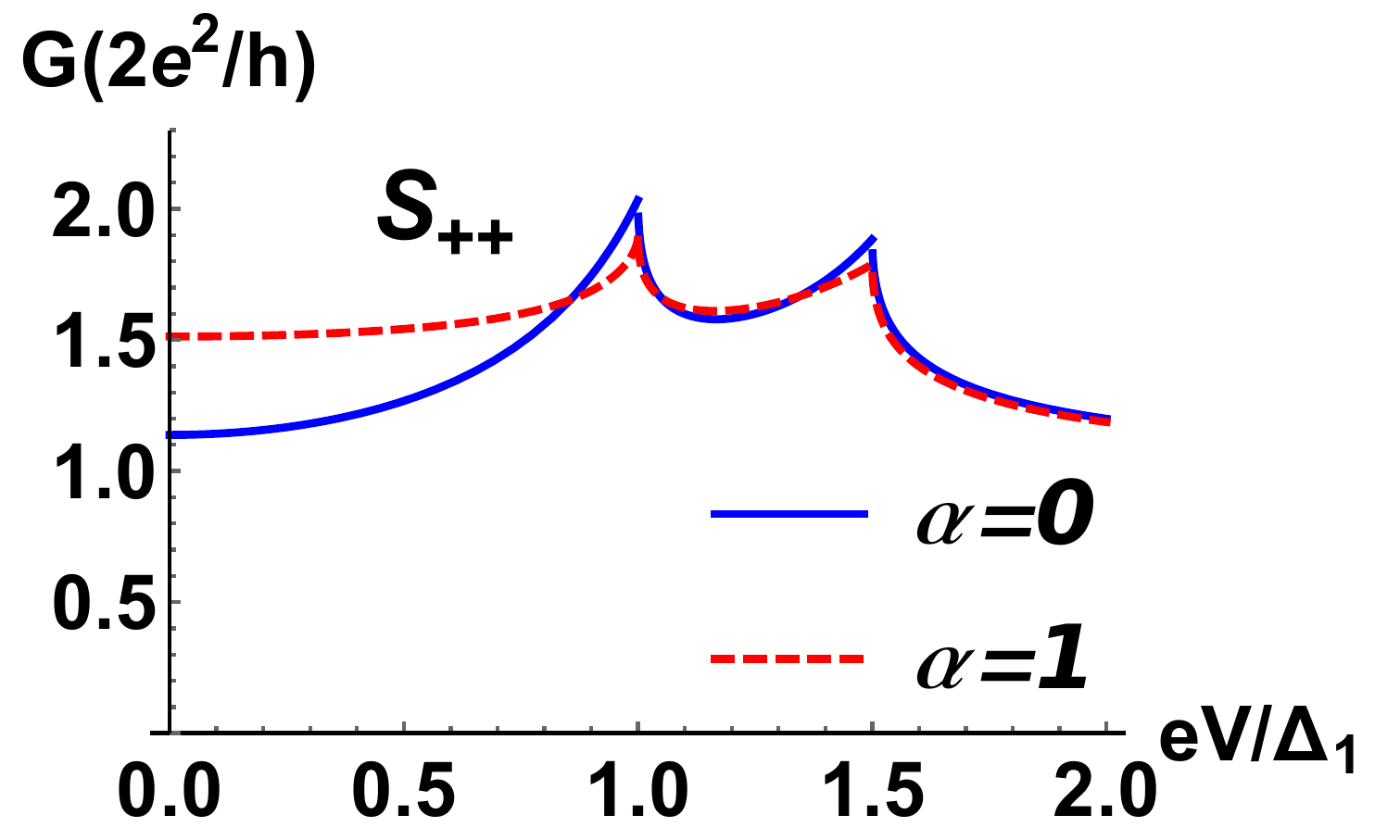}}
	\subfigure[]
	{\includegraphics[scale=0.478]{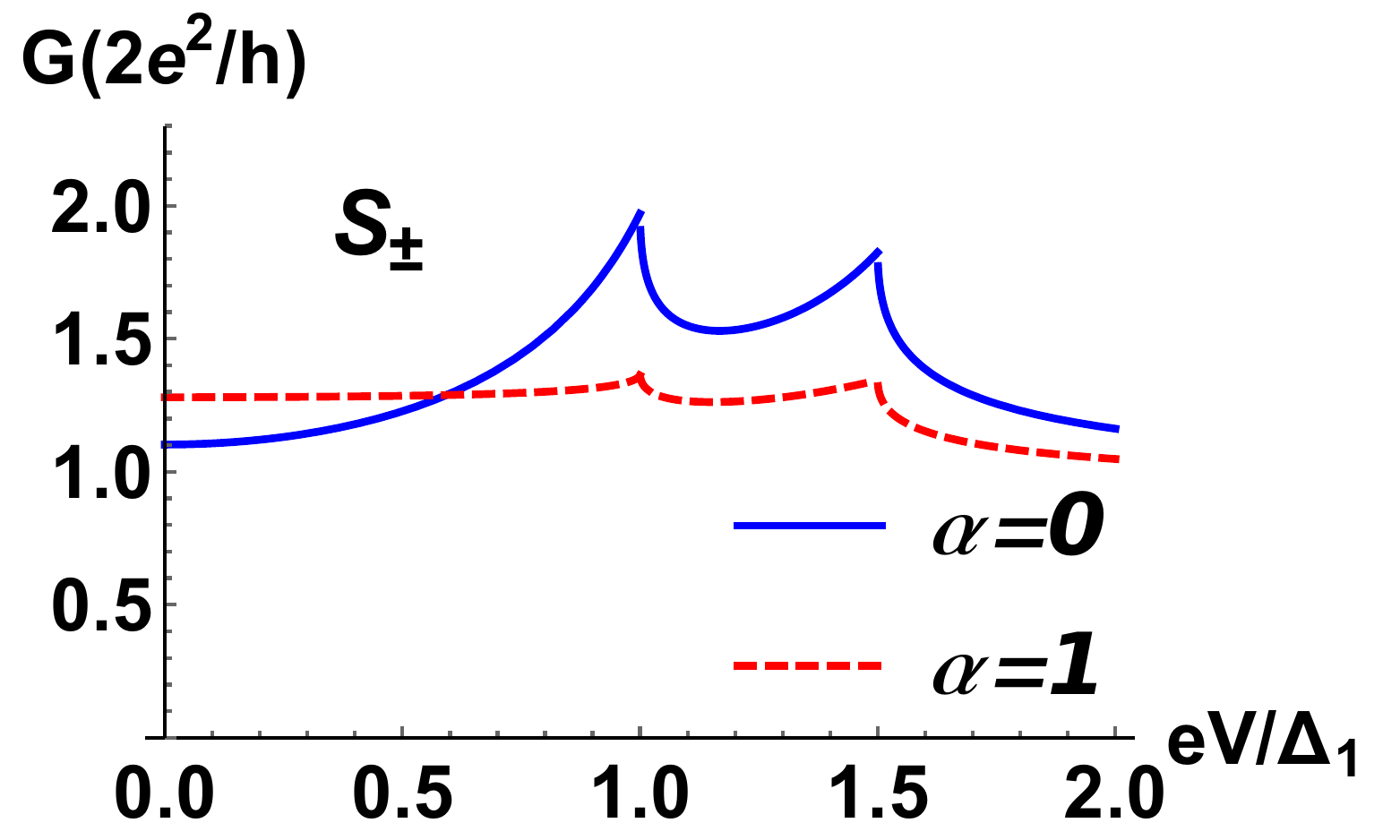}	}
    \caption{Differential conductance vs bias voltage($E/ \Delta_1$) for $\alpha=0.0$ and $1.0$, with $z=1.0$ for (a) $s_{++}$ pairing (b) $s_{\pm}$ pairing for Normal metal/Insulator/Normal metal/Insulator/Iron pnictide ($N_{1}/I/N_{2}/I/Ip$) superconductor junction with $a=0$. These are identical to that seen in Fig.~3 of Ref.~[\cite{linder}] for a $N/I/Ip$ junction.}
    \label{1}
\end{figure}
\begin{figure}[h]
\hspace{-0.61cm}\subfigure[]
	{\includegraphics[scale=0.46]{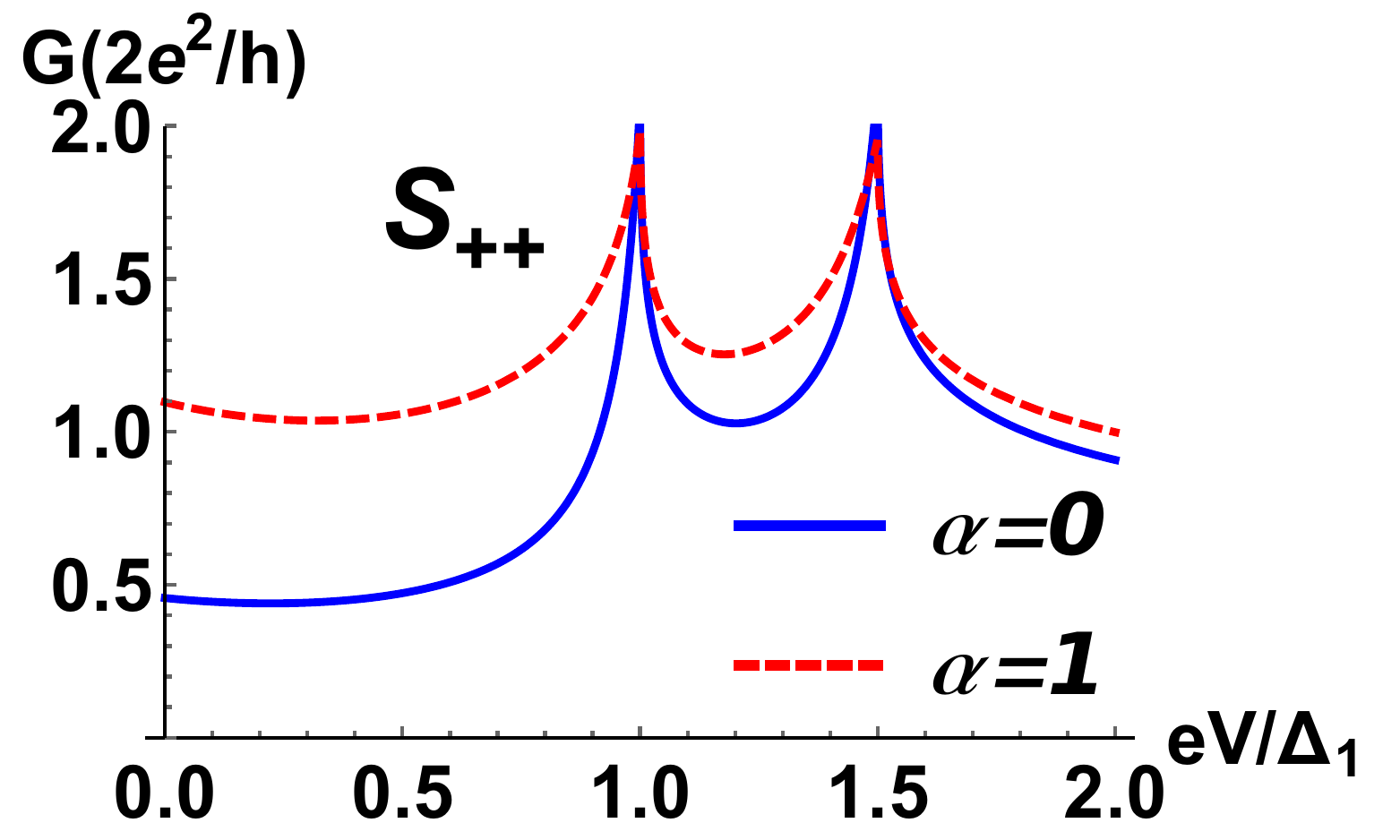}}
	\subfigure[]
	{\includegraphics[scale=0.46]{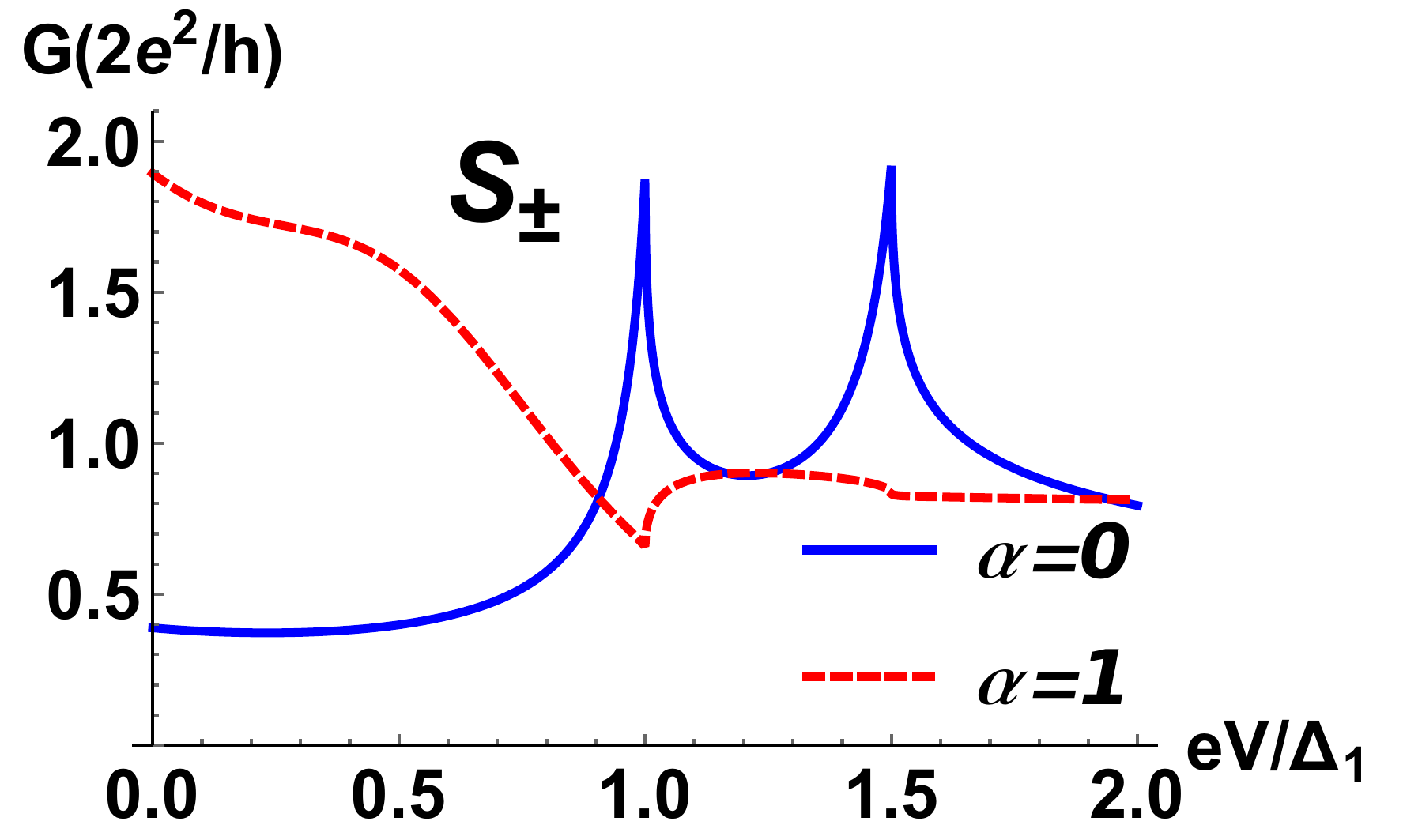}	}
	\caption{Differential conductance vs bias voltage($E/ \Delta_1$) with $\alpha=0.0$ and $1.0$,  $a=0.1 nm$ and $z_{1}=z_{2}=1.0$ for (a) $s_{++}$ pairing (b) $s_{\pm}$ pairing for $N_1/I/N_2/I/Ip$ superconductor junction.}
	\label{2}
\end{figure}
\subsection{Comparing differential conductance and differential Shot Noise for $N_1/I/N_2/I/Ip$ with $N/I/Ip$ junction}
We plot the differential conductance in the limit of vanishing length of second normal metal $N_2$($a \rightarrow 0$) as shown in Figs.~2(a) and 2(b) which are identical to that seen in Figs.~3(a) and 3(b) of Ref.~\cite{linder}, where they plot the differential conductance in a $N/I/Ip$ superconductor junction.  As can be clearly seen regardless of interband coupling strength($\alpha$) there is no zero bias conductance peak observed for either $s_{\pm}$ or $s_{++}$ pairing symmetries in a Normal metal/Insulator/Iron pnictide($N/I/Ip$) superconductor junction. In Fig.~3(a) and Fig.~3(b), we plot the differential conductance for a  $N_1/I/N_2/I/Ip$ superconductor junction with thickness of intermediate layer($a=0.1 nm$) and one can clearly see a ZBCP in case of $s_{\pm}$ pairing for $\alpha=1$ unlike the $N/I/Ip$ junction. 
\begin{figure}[t]
\hspace{-0.65cm}
        \subfigure[]
{    \includegraphics[scale=0.473]{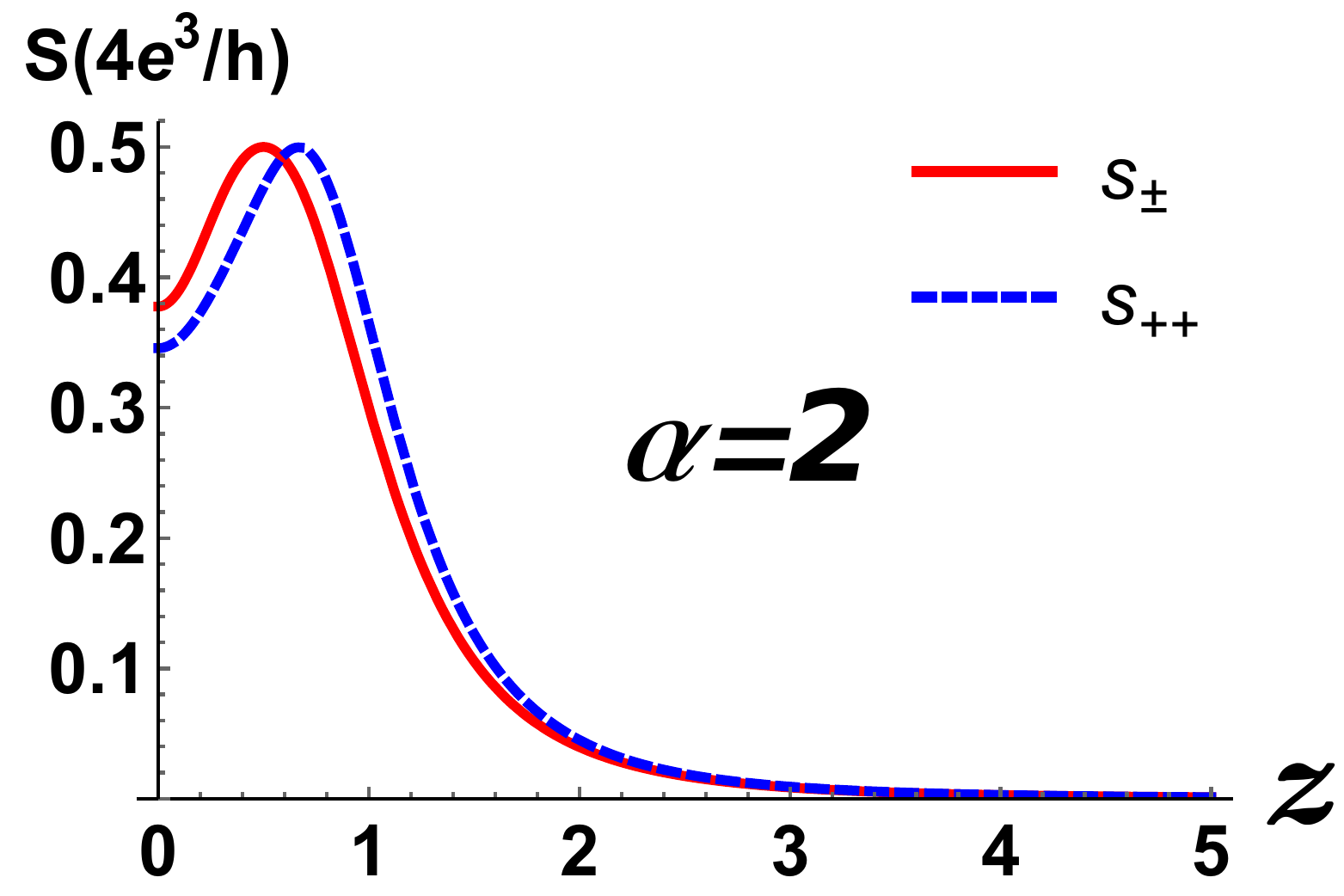} }  
        \subfigure[]
   { \includegraphics[scale=0.473]{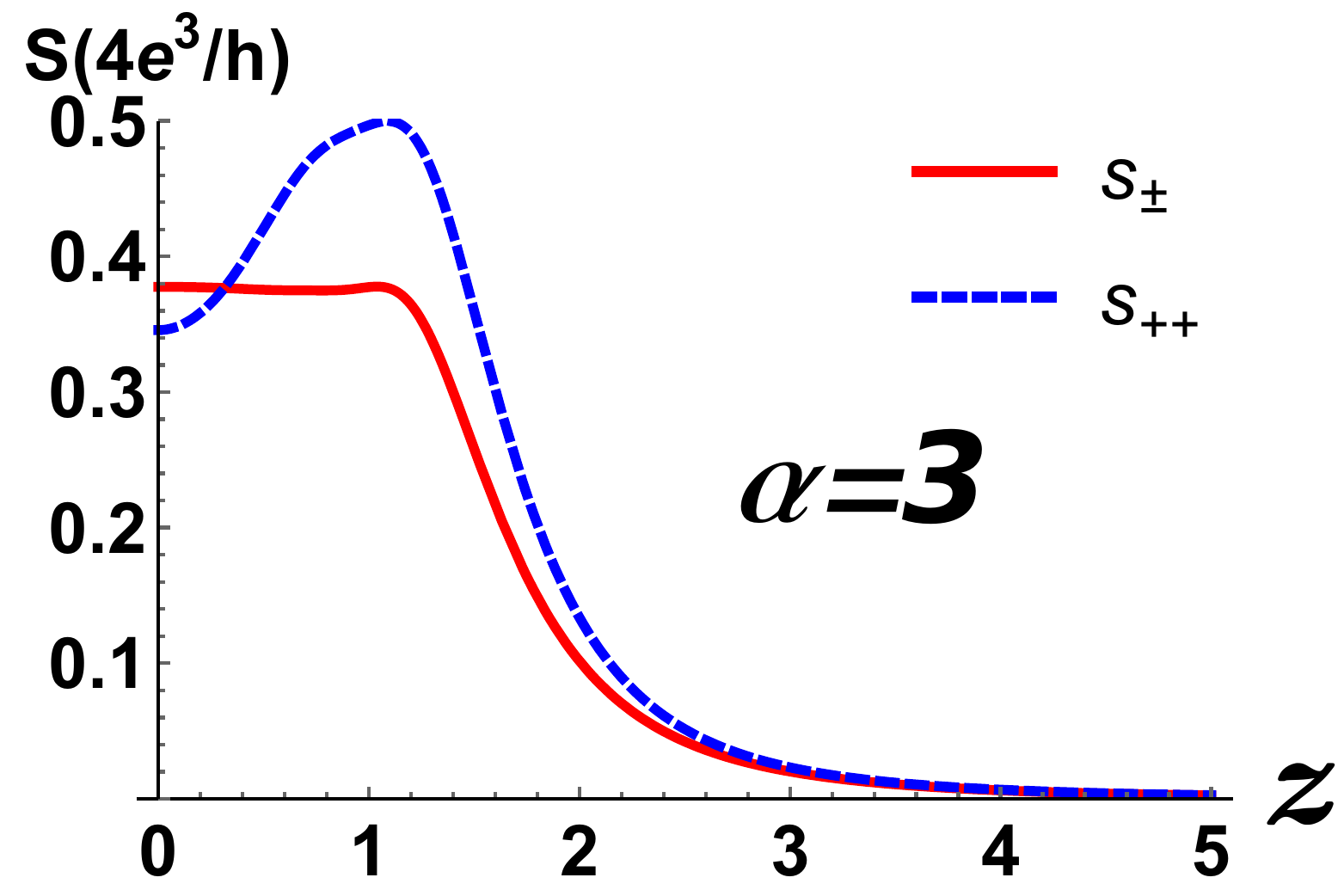}}
  \caption{Differential shot noise vs barrier strength ($z$) with $a= 0$, $z_1=z_2=z$ and  $E/ \Delta_1=1$ for (a) $\alpha=2$  (b) $\alpha=3$  for $N/I/Ip$ superconductor junction.}
\end{figure}
From Fig.~4, it is seen that regardless of barrier strength($z$) shot noise for both s$_\pm$ and s$_{++}$ pairing tend to zero for a $N/I/Ip$ superconductor junction in the tunnel limit even for different values of interband coupling(for example $\alpha = 2$ and $\alpha = 3$). On the contrary in Fig.~4 {\em of the main manuscript} in the tunnel limit ($z\rightarrow large$) shot noise for $s_{\pm}$ pairing tends to zero while shot noise for $s_{++}$ pairing tends to finite value for large interband coupling strength ($\alpha=2$ and $\alpha=3$) in a $N_1/I/N_2/I/Ip$ superconductor junction. This again shows the necessity of the non-superconducting bilayer in discriminating between the pairing symmetries which is not possible with a single non-superconducting layer.
\begin{figure}[h]
\hspace{-0.75cm}
        \subfigure[]
 {   \includegraphics[scale=0.46]{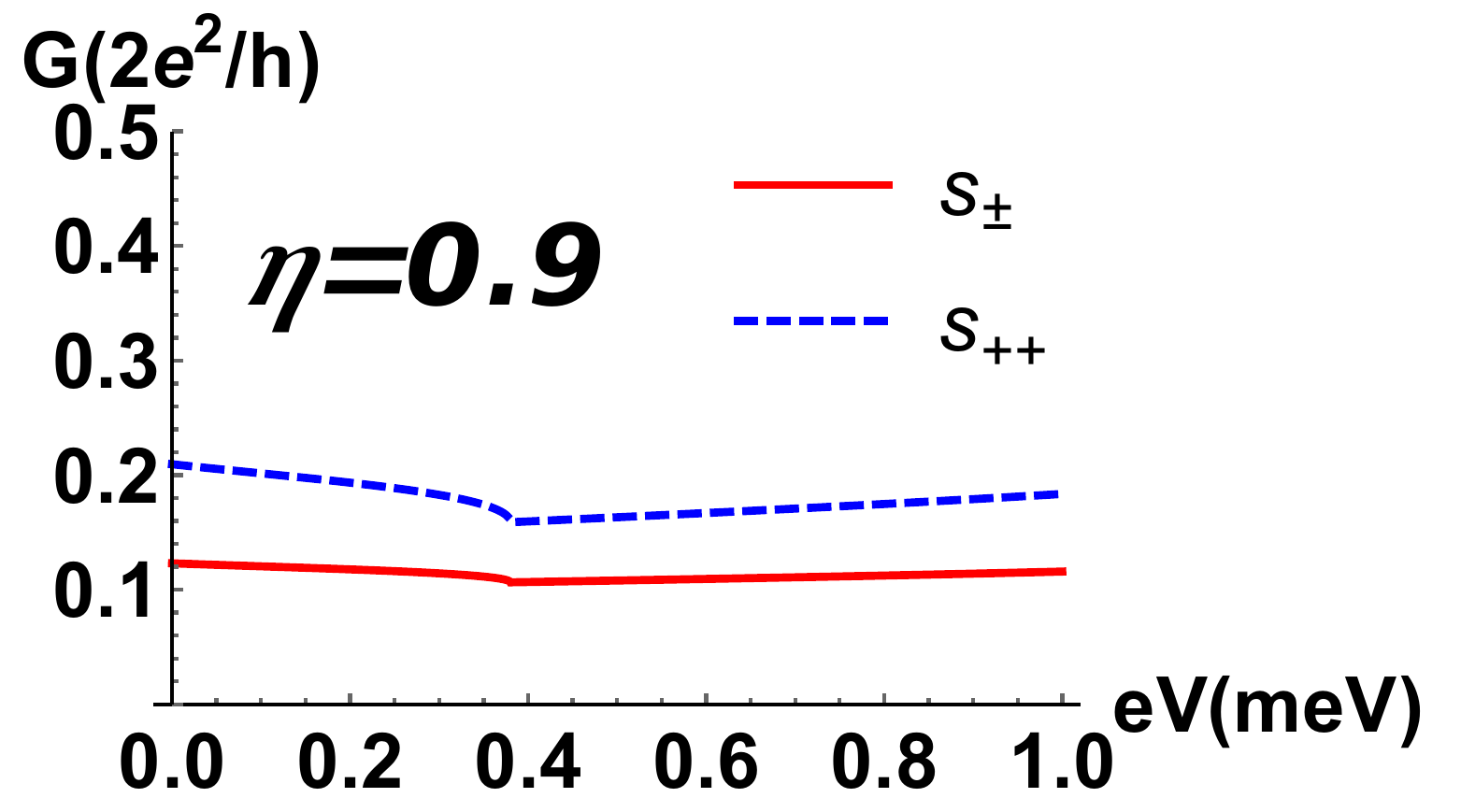} }
        \subfigure[]
   { \includegraphics[scale=0.46]{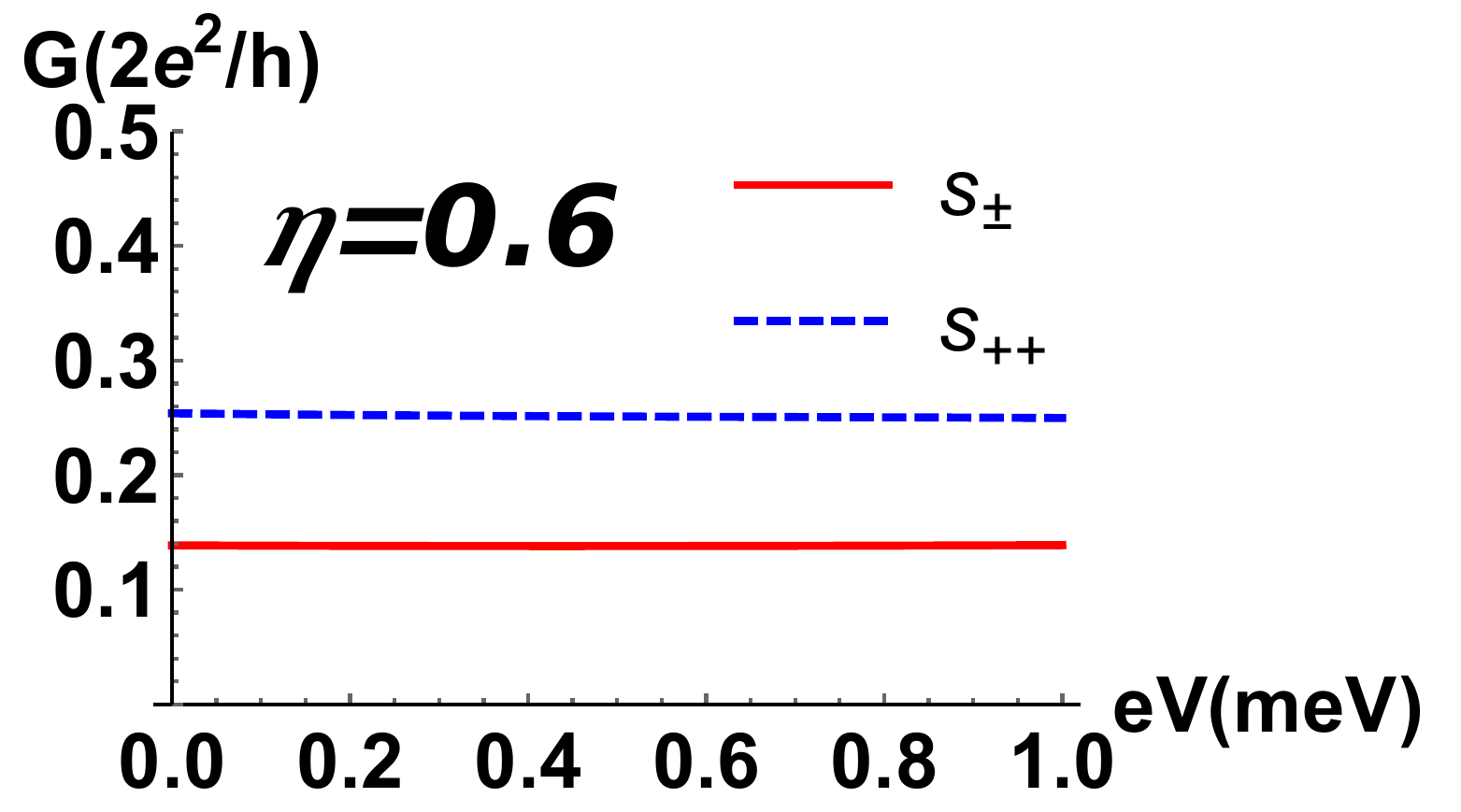}}
  \caption{Differential conductance for a $FM/I/Ip$ superconductor junction vs bias voltage V(meV) with $\Delta_2$=3.75meV, $\Delta_1$=2.5meV, $E_F$=3.8eV, $a=0$, $z_2$=$z_1=2$ and $\alpha=2$ for (a) $\eta=0.9$, (b) $\eta=0.6$.}
\end{figure}
\subsection{Comparing Differential conductance oscillation of $FM/I/NM/I/Ip$ with $FM/I/Ip$ junction}
We plot differential conductance for a $FM/I/Ip$ superconductor junction as seen in Fig.~5 for different magnetization values (for example $\eta=0.9$ and $\eta=0.6$). In Fig.~5 it is seen that regardless of any change in magnetization there is no differential conductance oscillation for a $FM/I/Ip$ superconductor junction in s$_{++}$ pairing nor in s$_{\pm}$ pairing. On contrary to there is differential conductance oscillation for a $FM/I/NM/I/Ip$ superconductor junction as seen in Fig.~7 {\em of main manuscript}. For a $FM/I/NM/I/Ip$ superconductor junction, the period of conductance oscillation for s$_{++}$ pairing is half the period of s$_{\pm}$ pairing which can be very helpful to probe the pairing symmetry of iron pnictide.

\end{document}